\global\def\draftcontrol{0}

   \def\versionno{ susysin}
\catcode`\@=11

\expandafter\ifx\csname draftcontrol\endcsname\relax\global\def\draftcontrol{0}
\fi

{\count255=\time\divide\count255 by 60
\xdef\hourmin{\number\count255}
\multiply\count255 by-60\advance\count255 by\time
\xdef\hourmin{\hourmin:\ifnum\count255<10 0\fi\the\count255}}
\def\draftdate{\number\month/\number\day/\number\year\ \ \ \hourmin }

\newcommand\makepapertitle{\par
  \begingroup
    \renewcommand\thefootnote{\@fnsymbol\c@footnote}%
    \def\@makefnmark{\rlap{\@textsuperscript{\normalfont\@thefnmark}}}%
    \long\def\@makefntext##1{\parindent 1em\noindent
            \hb@xt@1.8em{%
                \hss\@textsuperscript{\normalfont\@thefnmark}}##1}%
     \newpage
     \global\@topnum\z@   % Prevents figures from going at top of page.
     \@makepapertitle
     \thispagestyle{empty}\@thanks
  \endgroup
  \setcounter{footnote}{0}%
  \global\let\thanks\relax
  \global\let\makepapertitle\relax
  \global\let\@makepapertitle\relax
  \global\let\@thanks\@empty
  \global\let\@author\@empty
  \global\let\@date\@empty
  \global\let\@title\@empty
  \global\let\title\relax
  \global\let\author\relax
  \global\let\date\relax
  \global\let\and\relax
  \def\version{\let\version\@version\@gobble}
}
\def\@makepapertitle{%
  \newpage
   \ifnum\draftcontrol=1 {}
   \version\versionno
   \vskip 3em%
   \else
   \hfill\hbox to 3cm {\parbox{4cm}{\@pubnum}\hss}%
   \vskip 3em%
   \fi
   \begin{center}%
   \let \footnote \thanks
     {\LARGE {\@title}}%
     \vskip 1.5em%
     {\normalsize%\large
       \lineskip .5em%
       \begin{tabular}[t]{c}%
         \@author
       \end{tabular}\par}%
     \vskip 1.5em%
     {\@bstract}%
     \end{center}%
     \vskip 1.5em
     \@date%
   \par
}

\gdef\@pubnum{}
\def\pubnum#1{%
  \gdef\@pubnum{#1}}

\gdef\@bstract{}
\def\Abstract#1{%
  \gdef\@bstract{%
   \parbox{\textwidth-0pc}{%
   \centerline{\bf Abstract}\penalty1000%
\kern.2cm%
\noindent%\abstractfont \baselineskip=12pt
\renewcommand\baselinestretch{1.0}%
{#1}}}
}

\def\ps@paper{\let\@mkboth\@gobbletwo%
     \ifnum\draftcontrol=1
    \def\@oddfoot{\hbox to \textwidth{\tiny \versionno \hfil\tiny\draftdate}%
    \hskip -\textwidth \hbox to \textwidth{\hfil\rm\thepage\hfil}}%
     \else\def\@oddfoot{\hbox to \textwidth{\hfil\rm\thepage\hfil}}
     \fi
     \let\@evenfoot\@oddfoot
}
\def\body{\clearpage
          \pagestyle{paper}
    }
\def\@version#1{\ifnum\draftcontrol=1
\typeout{}\typeout{#1}\typeout{}
\vskip3mm\centerline{\hbox{\fbox{\normalsize{\tt DRAFT -- #1 -- }
                   {\draftdate}}}}\vskip3mm
\fi}
\let\version\@version
\long\def\eqlabel#1{\ifnum\draftcontrol=1
                    \tag@false  % there are some problems with multline without this
                    \tag*{(\theequation) \hbox to -0.2cm{\hspace{0cm}\small{#1}\hss}}
                    \refstepcounter{equation}
                    \edef\@currentlabel{\theequation}
                    \ltx@label{#1}          % use old LaTeX \label instead of new definition
                    \else
                    \label{#1}
                    \fi
                    }
\let\st@bibitem\@bibitem
\let\st@lbibitem\@lbibitem
\ifnum\draftcontrol=1
  \def\@bibitem#1{%
    \st@bibitem{#1}\a@@label{#1}\ignorespaces}
  \def\@lbibitem[#1]#2{%
    \st@lbibitem[#1]{#2}\a@@label{#2}\ignorespaces}
  \def\a@@label#1{%
    \gdef\a@lab{\smash{\normalfont\small#1}}
    \ifvmode
      \if@inlabel
        \global\setbox\@labels\hbox{%
          \llap{\a@lab\let\a@lab\relax
                \kern\@totalleftmargin\kern\marginparsep}%
          \box\@labels}%
      \fi
    \fi}
\fi
\documentclass[12pt,letterpaper]{article}
\usepackage{amsmath,amssymb,array,calc,epsfig,rotating,bm}
\usepackage[sort]{cite}
\usepackage{graphicx}
\usepackage{psfrag,verbatim}

\ifnum\draftcontrol=1
\fi
\renewcommand\baselinestretch{1.25}
\renewcommand\section{\@startsection {section}{1}{\z@}%
                                   {-3.5ex \@plus -1ex \@minus -.2ex}%
                                   {2.3ex \@plus.2ex}%
                                   {\normalfont\large\bfseries}}
\renewcommand\subsection{\@startsection{subsection}{2}{\z@}%
                                   {-3.25ex\@plus -1ex \@minus -.2ex}%
                                   {1.5ex \@plus .2ex}%
                                   {\normalfont\normalsize\bfseries}}
\renewcommand\subsubsection{\@startsection{subsubsection}{3}{\z@}%
                                   {-3.25ex\@plus -1ex \@minus -.2ex}%
                                   {1.5ex \@plus .2ex}%
                                   {\normalfont\normalsize\it}}
\renewcommand\paragraph{\@startsection{paragraph}{4}{\z@}%
                                   {-3.25ex\@plus -1ex \@minus -.2ex}%
                                   {1.5ex \@plus .2ex}%
                                   {\normalfont\normalsize\bf}}

\numberwithin{equation}{section}

\def\revise#1       {\raisebox{-0em}{\rule{3pt}{1em}}%
                     \marginpar{\raisebox{.5em}{\vrule width3pt\
                     \vrule width0pt height 0pt depth0.5em
                     \hbox to 0cm{\hspace{0cm}{%
                     \parbox[t]{4em}{\raggedright\footnotesize{#1}}}\hss}}}}

\newcommand\nxt[1]  {\\\fnxt#1}
\newcommand{\ie}{{\it i.e.,}\ }
\newcommand{\eg}{{\it e.g.,}\ }

\def\cale         {{\cal E}}

\def\call         {{\cal L}}
\def\calm         {{\cal M}}
\def\caln         {{\cal N}}
\def\calo         {{\cal O}}
\def\calp         {{\cal P}}

\def\calw         {{\cal W}}

\def\zet          {{\mathbb Z}}

\def\del          {\partial}

\def\Re           {{\rm Re\hskip0.1em}}
\def\Im           {{\rm Im\hskip0.1em}}

\def\sqr#1#2{{\vcenter{\vbox{\hrule height.#2pt
 \hbox{\vrule width.#2pt height#1pt \kern#1pt
 \vrule width.#2pt}\hrule height.#2pt}}}}

\def\b{\beta}
\def\w{\omega}
\def\r{\rho}
\def\dd{\delta}

\def\aa1{\phi}
\def\cc1{\psi}

\def\k{\kappa}

\def\l{\lambda}

\def\k{\kappa}

\def\ds{d\sigma}

\catcode`\@=12

\begin{document}

%%%
%%%%%% text starts here
%%%%%%%%%

\title{\bf Singularity development and supersymmetry in holography}

\date{May 23, 2017}
%\date\today

\author{
Alex Buchel\\[0.4cm]
\it $ $Department of Applied Mathematics\\
\it $ $Department of Physics and Astronomy\\ 
\it University of Western Ontario\\
\it London, Ontario N6A 5B7, Canada\\
\it $ $Perimeter Institute for Theoretical Physics\\
\it Waterloo, Ontario N2J 2W9, Canada
}

\Abstract{We study the effects of supersymmetry on singularity development
scenario in holography presented in \cite{Bosch:2017ccw} (BBL). We
argue that the singularity persists in a supersymmetric extension of
the BBL model. The challenge remains to find a string theory embedding
of the singularity mechanism.
}

\makepapertitle

\body

\version\versionno
\tableofcontents

\section{Introduction}\label{intro}

Horizons are ubiquitous in holographic gauge theory/string theory correspondence \cite{m1,Aharony:1999ti}.
Static horizons are dual to thermal states of the boundary gauge theory  \cite{Witten:1998zw}, 
while their long-wavelength  near-equilibrium dynamics encode the effective boundary hydrodynamics of the 
theory \cite{Bhattacharyya:2008jc}. Typically, dissipative effects in the hydrodynamics (due to shear and bulk viscosities)
lead to an equilibration of a gauge theory state --- a slightly perturbed horizon in a dual 
gravitational description settles to an equilibrium configuration. If the initial perturbation away from thermal equilibrium 
is sufficiently strong, non-hydrodynamic modes participate in the equilibration 
process\footnote{Even when not excited, non-hydrodynamic modes are important as they determine 
the convergence properties of the all-order effective hydrodynamic description 
\cite{Heller:2013fn,Buchel:2016cbj}. } --- perturbed bulk horizon relaxes via quasinormal 
modes \cite{Berti:2009kk,Buchel:2015saa,Fuini:2015hba,Janik:2015waa,Buchel:2015ofa}. 
We call horizons that relax via hydrodynamic modes that attenuate in space-time domain  
or via positively gapped quasinormal modes\footnote{{\it I.e}, the quasinormal modes with $\Im(\omega)<0$.}   
{\it stable horizons}. Bulk holographic dual with (dynamically) stable horizons describe boundary 
gauge theory states which are stable with respect to sufficiently small fluctuations. 

Translationary invariant horizon can suffer an instability when a hydrodynamic mode 
in a system becomes unstable. An example of such an instability is discussed in 
\cite{Buchel:2005nt}. In this case, the expected end-point of the evolution is a new 
inhomogeneous phase of the system\footnote{See \cite{Attems:2017ezz} for a recent discussion.}.  
Alternatively, a horizon can be de-stabilized when a positively gapped non-hydrodynamic mode 
becomes unstable when one lowers energy (or temperature) \cite{Hartnoll:2008vx,Hartnoll:2008kx}.
This instability realizes a holographic dual of a spontaneous symmetry breaking in a 
mean-field approximation --- the system undergoes a second-order phase transition 
towards a new stable phase with a finite density condensate of the originally unstable mode. 
  
Above classification of (un)stable horizons in a holographic framework matches both  
the gravitational and the field theory intuition. However, there is more to the story.
In  \cite{Buchel:2009ge,Buchel:2010wk} a new instability of the translationary invariant 
holographic horizons was identified: 
\nxt while there is a linearized instability below some critical energy density 
(or temperature), triggered by a non-hydrodynamic mode spontaneously breaking a discrete symmetry,
\nxt there is no candidate equilibrium state with a condensate of this unstable mode.\\
The dynamics of the model \cite{Buchel:2009ge,Buchel:2010wk} was extensively studied in 
\cite{Bosch:2017ccw} (BBL): both the presence of the linearized instability and the 
absence of the suitable equilibrium state for the evolution below the criticality 
was confirmed dynamically. Moreover, it was argued that the gravitational system evolves 
to a region of arbitrary large curvatures in the vicinity of the horizon, asymptotically turning 
such region singular in a finite time with respect to the boundary theory. The area density of the apparent 
horizon, associated with the non-equilibrium entropy density of the boundary gauge theory 
\cite{Booth:2005qc,Figueras:2009iu} also diverges within a finite boundary 
time. This latter observation is significant as it precludes evolution of instabilities towards any 
finite entropy density spatially inhomogeneous equilibrium states as well --- in other words, despite the 
fact that BBL dynamics occurs in a constrained phase space (spatial homogeneity and isotropy), the conclusion 
that the system evolves to a singularity, violating the weak cosmic censorship conjecture, 
is robust. 

The BBL model is not a top-down holographic construction. Thus, one might wonder whether the phenomenon discovered 
in  \cite{Buchel:2009ge,Buchel:2010wk,Bosch:2017ccw} occurs in real string theory holographic examples. 
A particular aspect of the model is the unboundedness of the bulk scalar field potential. It was argued in 
\cite{Bosch:2017ccw} that there is no weak cosmic censorship conjecture violation once the potential is bounded. 
On the other hand, unbounded scalar potentials do occur in supersymmetric top-down holographic models (as \eg in 
\cite{Donos:2011ut}). 
Additionally, the exotic critical phenomena of  \cite{Buchel:2009ge} was identified in supergravity 
model \cite{Donos:2011ut} (DG) in grand canonical ensemble. In this paper 
we partly address above questions.  

Unfortunately, construction of a top-down holographic model realizing the dynamics of BBL model 
remains open: while the DG model is "exotic'' in grand canonical ensemble, we study in 
section \ref{dg} the equilibrium properties and dynamics of DG model in microcanonical ensemble  
and show that it realizes a standard spontaneous symmetry breaking instability as in  
\cite{Hartnoll:2008vx,Hartnoll:2008kx}. In section \ref{bblsusy} we present a supersymmetric 
extension of the BBL model, which exhibits the exotic phenomenon of \cite{Buchel:2009ge,Buchel:2010wk}. 
This sBBL model does not have an equilibrium state below the criticality, and, as in BBL model, 
its homogeneous and isotropic states  evolve
towards asymptotically divergent expectation value of the symmetry breaking operator.
The similarities and differences of the BBL and sBBL model are further highlighted in 
section \ref{conclude}. Some technical details are delegated to appendix \ref{appendix}.

\section{DG model in microcanonical ensemble}\label{dg}

In \cite{Donos:2011ut} the authors studied equilibrium states of $d=3$ $\caln=8$ superconformal 
gauge theory dual to $AdS_4\times S^7$ at a finite temperature and a finite chemical potential with respect to a diagonal 
$U(1)_R\subset SO(8)$ global symmetry within different consistent truncations of $D=11$ supergravity on $S^7$,
allowing for  different patterns of spontaneous global symmetry breaking. We consider the following two consistent 
truncations, with the effective actions
\begin{equation}
S_{DG-A/B} = \int_{\calm_4} dx^4\sqrt{-\gamma}\ \call_{DG-A/B}\,,  
\eqlabel{dgaction}
\end{equation}
where 
\begin{itemize}
\item ``DG-A'' model Lagrangian ({\it $2+2$ equal charged scalars} model in \cite{Donos:2011ut}) is :
\begin{equation}
\begin{split}
&\call_{DG-A}=\frac 12 R -\frac 14 \left(\del\sigma\right)^2-\frac 12 \left(\del \gamma_1\right)^2
-\frac 12 \left(\del \gamma_2\right)^2-\frac 12\sinh^2\gamma_1 \left(A^1+\bar{A}^0\right)^2\\
&-\frac 12\sinh^2\gamma_2 \left(A^1-\bar{A}^0\right)^2-\frac14\cosh(\sigma)\left[
\left(\bar{F}^0\right)_{\mu\nu}\left(\bar{F}^0\right)^{\mu\nu}+\left({F}^1\right)_{\mu\nu}\left({F}^1\right)^{\mu\nu}
\right]\\
&-\frac 12 \sinh(\sigma)\left[
\left({F}^1\right)_{\mu\nu}\left(\bar{F}^0\right)^{\mu\nu}\right]-V\,,
\end{split}
\eqlabel{dgla}
\end{equation}
with 
\begin{equation}
V=-2\ \left(2\ \cosh\gamma_1\ \cosh\gamma_2+\cosh\sigma\right)\,;
\eqlabel{va}
\end{equation}
\item  "DG-B'' model Lagrangian ({\it $4$ equal charged scalars} model in \cite{Donos:2011ut}) is:
\begin{equation}
\begin{split}
&\call_{DG-B}=\frac 12 R -\left(\del \phi\right)^2-\sinh^2\phi \left(A^1\right)^2
-\frac14\ 
\left({F}^1\right)_{\mu\nu}\left({F}^1\right)^{\mu\nu}-V\,,
\end{split}
\eqlabel{dglb}
\end{equation}
with 
\begin{equation}
V=-2\ \left(2+ \cosh(2\phi)\right)\,.
\eqlabel{vb}
\end{equation}
\end{itemize}  
For the maximally supersymmetric quantization the dimensions $\Delta$ of the CFT 
operators $\calo$ dual to bulk scalars $\{\gamma_1,\gamma_2,\sigma\}$ (DG-A model) are 
\begin{equation}
\Delta(\calo_{\gamma_1})=\Delta(\calo_{\gamma_2})=\Delta(\calo_{\sigma})=1\,,
\eqlabel{dima}
\end{equation}
and to the bulk scalar $\phi$ (DG-B model) is  
\begin{equation}
\Delta(\calo_{\phi})=1\,.
\eqlabel{dimb}
\end{equation}
In both models $A^1$ is the bulk gauge field dual to $U(1)_R$ global symmetry; 
the bulk scalars $\gamma_i$ and $\phi$  have the $U(1)_R$ charge 1, while $\sigma$ is $R$-symmetry neutral. 

Note that DG-A model is invariant under the $\zet_2$ symmetry:
\begin{equation}
\bar{A}^0\to -\bar{A}^0\,,\qquad \sigma\to -\sigma\,,\qquad \gamma_1\leftrightarrow \gamma_2\,.
\eqlabel{daz2}
\end{equation}
Its truncation to a $\zet_2$-even sector produces DG-B model with the identification 
\begin{equation}
\gamma_1= \gamma_2\ \equiv\ \phi\,.
\eqlabel{dadb}
\end{equation}

\subsection{Grand canonical ensemble (a review)}

We focus on DG-A model; the construction for the DG-B model is a consistent truncation as explained in \eqref{dadb}.
Following \cite{Donos:2011ut}, we set the chemical potential $\mu_1=1$ for $U(1)_R$ symmetry, and for the other 
global $U(1)$ (holographic dual to the bulk gauge field $\bar{A}^0$) $\bar{\mu}^0=0$, \ie spatially homogeneous and 
isotropic thermal equilibrium states of the 
CFT are represented by a bulk geometry with the asymptotics:  
\begin{equation}
\begin{split}
&A^1=a_1(r)\ dt\,,\qquad a_1=\mu_1+\frac{q_1}{r}+\calo(r^{-2})\,,\\
&\bar{A}^0=\bar{a}_0(r)\ dt\,,\qquad \bar{a}_0=\frac{\bar{q}_0}{r}+\calo(r^{-2})\,,
\end{split}
\eqlabel{gauge}
\end{equation}
where $r$ is a standard asymptotic-AdS radial coordinate (see \eqref{metric} below) and $\{q_1,\bar{q}_0\}$ determine the $U(1)_R\times U(1)$ 
charge densities, see \eqref{thprop}. Imposing the supersymmetric quantization \eqref{dima}, 
\begin{equation}
\gamma_i=\frac{\gamma_{i(1)}}{r}+\calo(r^{-3})\,,\qquad \sigma=\frac{\sigma_{(1)}}{r}+\calo(r^{-3})\,,
\eqlabel{vev}
\end{equation} 
we identify the expectation values of the dual operators 
\begin{equation}
\langle \calo_{\gamma_i}\rangle = \gamma_{i(1)}\,,\qquad \langle \calo_{\sigma}\rangle =\sigma_{(1)}\,.
\eqlabel{vevs}
\end{equation}
The background metric ansatz takes form 
\begin{equation}
ds_4^2=-2 r^2 e^{-\beta(r)}g(r)\ dt^2 +\frac{dr^2}{2r^2 g(r)}+r^2 \left(dx_1^2+dx_2^2\right)\,,
\eqlabel{metric}
\end{equation}
with the following asymptotic expansions at infinity:
\begin{equation}
\begin{split}
&g=1+\frac{1}{2r^2}\left(\gamma_{1(1)}^2+\gamma_{1(2)}^2+\frac12 \sigma_{(1)}^2\right)+\frac{m}{r^3}+\calo(r^{-4})\,,\\
&\b=\frac{1}{2r^2}\left(\gamma_{1(1)}^2+\gamma_{1(2)}^2+\frac12 \sigma_{(1)}^2\right)+\calo(r^{-4})\,,
\end{split}
\eqlabel{metuv}
\end{equation}
where parameter $m$ is related to the energy density of the state, see \eqref{thprop}.
Assuming that a regular Schwarzschild horizon is located at $r=r_h$, we have asymptotic expansions 
in $(r-r_h)>0$ as 
\begin{equation}
\begin{split}
&g=g_h^{(1)} (r-r_h)+\cdots\,,\qquad \b=\b_h^{(0)}+\b_h^{(1)}(r-r_h)+\cdots\,,\\
&a_1=a^{(1)}_{1,h}(r-r_h)+\cdots\,,\qquad \bar{a}_0=\bar{a}^{(1)}_{0,h}(r-r_h)+\cdots\,,\\
&\gamma_i=\gamma_{i,h}^{(0)}+\gamma_{i,h}^{(1)}(r-r_h)+\cdots\,,\qquad \sigma=\sigma_h^{(0)}+\sigma_h^{(1)}(r-r_h)+\cdots\,.
\end{split}
\eqlabel{hor}
\end{equation}
Parameters in asymptotic expansions \eqref{gauge}, \eqref{vev}, \eqref{metuv} and \eqref{hor} 
determine the thermodynamic properties of an equilibrium CFT state, the energy density $\cale$, the pressure $P=\frac 12\cale$,
the grand potential density $\Omega=-P$, 
the temperature $T$, the entropy density $s$,  and the $U_R(1)\times U(1)$ charge density $\{\r_1,\bar{\r}_0\}$:
\begin{equation}
\begin{split}
&\cale=-2 m\,,\qquad T=\frac{r_h^2}{2\pi}\ g_h^{(1)} e^{-\b_h^{(0)}/2}\,,\qquad s=2\pi r_h^2\,,\qquad \r_1=-q_1\,,\qquad \bar{\r}_0=-\bar{q}_0\,.
\end{split}
\eqlabel{thprop}
\end{equation}

We use numerical shooting method developed in \cite{Aharony:2007vg} with a radial coordinate 
\begin{equation}
x\equiv \frac{r_h}{r}\,,\qquad x\in (0,1]\,,
\end{equation}
to relate the boundary and the near-horizon asymptotics of the dual gravitational background. We reproduce the results 
reported in  \cite{Donos:2011ut}:  
\nxt At any temperature $T>0$ there is a phase of the CFT with zero condensates\footnote{This is just an electrically charged $AdS_4$ RN black brane.} 
\eqref{vevs} and a grand potential density $\Omega_{RN}$:
\begin{equation}
T=\frac{12 r_h^2-\mu_1^2}{8\pi r_h}\,,\qquad \cale= 2 r_h^3 \left(1+\frac{\mu_1^2}{4r_h^2}\right)=-2 \Omega_{RN}\,,\qquad \r_1=\mu_1 r_h\,,
\eqlabel{rnads}
\end{equation} 
where $r_h\ge \mu_1/\sqrt{12}$.
\nxt There is a critical temperature $T_c$  in the system separating phases with 
nonzero condensates \eqref{vevs}  
\begin{equation}
\frac{T_c}{\mu_1}=0.1739106(2)\,,\qquad \frac{\r_1}{\mu_1^2}\bigg|_{T=T_c}=0.5234403(5)\,,\qquad \frac{\cale}{\mu_1^3}\bigg|_{T=T_c}=0.5485552(5)\,.
\eqlabel{tc}
\end{equation}
\nxt A phase with 
\begin{equation}
\langle \calo_{\gamma_1}\rangle = \langle \calo_{\gamma_2}\rangle \equiv \langle \calo_{\phi}\rangle\,,\qquad \langle \calo_{\sigma}\rangle =0\,,\qquad 
\bar{\r}_0=0\,,
\eqlabel{modelbphase}
\end{equation}
\ie model DG-B, exists at $T\ge T_c$. As its grand potential density $\Omega_{DG-B}$ is
\begin{equation}
\Omega_{DG-B}\ge \Omega_{RN} \bigg|_{\{T,\mu_1\}=const}\,,
\eqlabel{omb}
\end{equation}
DG-B model  realizes the exotic thermodynamics discovered in \cite{Buchel:2009ge} in grand canonical 
ensemble. 
\nxt A phase with 
\begin{equation}
\langle \calo_{\gamma_1}\rangle \ne \langle \calo_{\gamma_2}\rangle\ne 0\,,\qquad \langle \calo_{\sigma}\rangle \ne 0\,,\qquad 
\bar{\r}_0\ne 0\,,
\eqlabel{modelaphase}
\end{equation}
\ie model DG-A, exists at $T\le T_c$. As its grand potential density $\Omega_{DG-A}$ is
\begin{equation}
\Omega_{DG-A}\le \Omega_{RN} \bigg|_{\{T,\mu_1\}=const}\,,
\eqlabel{oma}
\end{equation}
DG-A model realized a standard holographic dual of a spontaneous symmetry breaking in grand canonical ensemble 
\cite{Hartnoll:2008vx,Hartnoll:2008kx}.

\begin{figure}[t]
\begin{center}
\psfrag{t}{{${T}/{\mu_1}$}}
\psfrag{w}{{${\Omega}/{\mu_1^3}$}}
\psfrag{v}{{$\langle\calo_{\gamma_1}\rangle$}}
\includegraphics[width=2.7in]{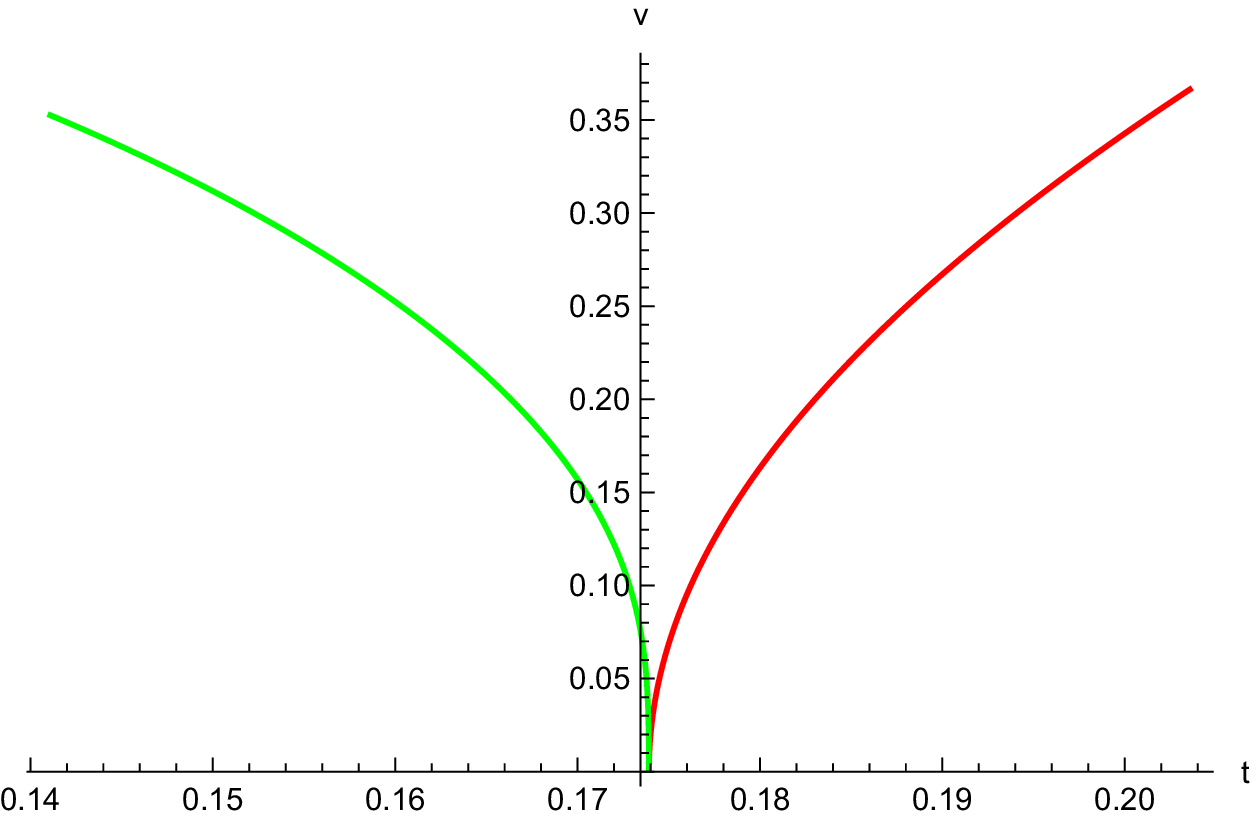}\qquad
\includegraphics[width=2.7in]{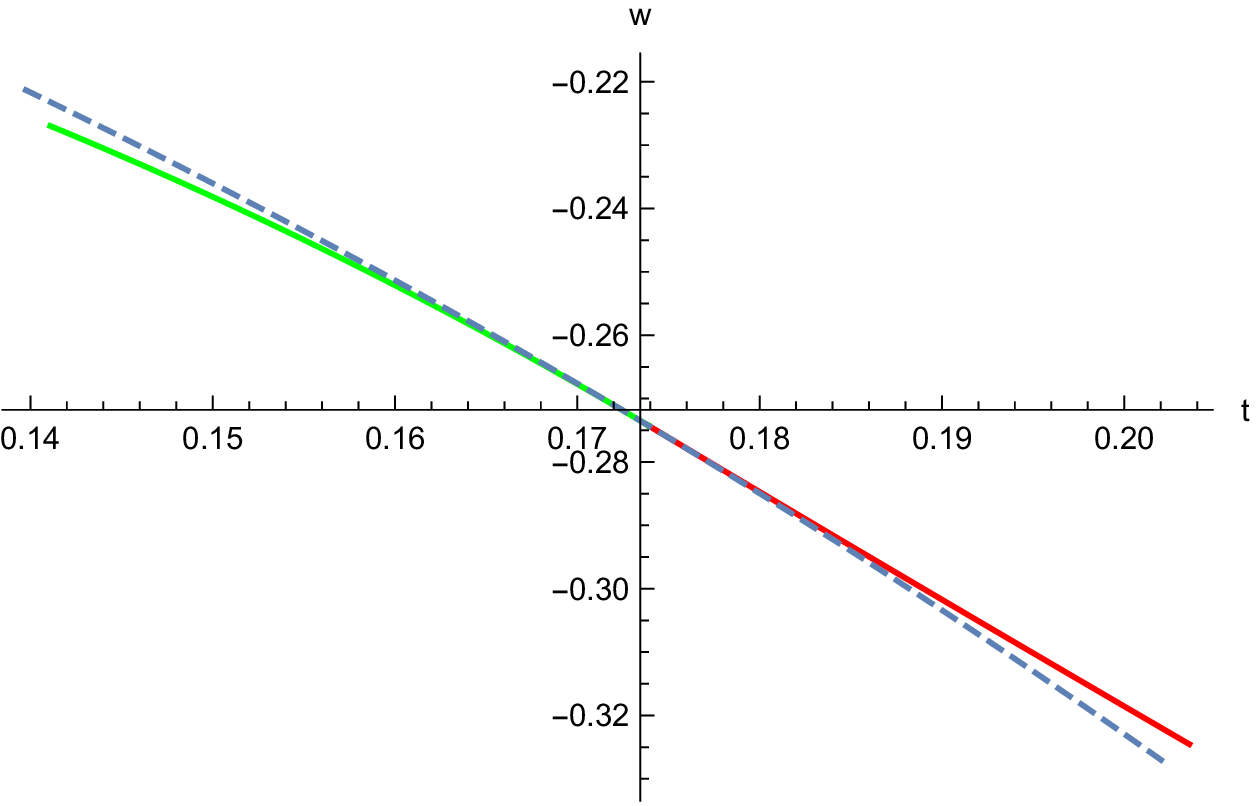}
\end{center}
  \caption{Left panel: expectation value $\langle\calo_{\gamma_1}\rangle$ as a function of $T/\mu_1$ in model 
DG-A (green curve) and model DG-B (red curve). Right panel: grand potential densities 
$\Omega=\{\Omega_{DG-A},\Omega_{DG-B},\Omega_{RN}\}=\{{\rm (green\ curve) }, 
{\rm (red\ curve) }, {\rm (dashed\ blue\ curve) }\}$  as a function of $T/\mu_1$. } \label{figure1}
\end{figure}

$\langle\calo_{\gamma_1}\rangle$ condensate in various phases of the CFT and the corresponding grand potential densities 
close to criticality, see \eqref{tc},  
are presented in fig.~\ref{figure1}.

\subsection{Microcanonical ensemble}\label{dgmicro}

In previous section we reproduced some of the results of \cite{Donos:2011ut} to confirm that model DG-B indeed exhibits 
exotic thermodynamics in the spirit of  \cite{Buchel:2009ge} in grand canonical ensemble.   
The singularity mechanism identified in \cite{Bosch:2017ccw} hinges upon the persistency of exotic phase structure in
microcanonical ensemble, \ie in dynamical evolution with fixed energy density and charges. 
Unfortunately, the instability mechanism  in  both  DG-A and DG-B models in microcanonical ensemble lead to 
standard picture of the spontaneous symmetry breaking\footnote{We verified that this remains 
true for supersymmetry breaking assignment of dimensions of bulk operators dual to $\{\gamma_i,\sigma\}$. } 
\cite{Hartnoll:2008vx,Hartnoll:2008kx}.

To study DG-A/B models in microcanonical ensemble, the only modification needed is the change of the 
boundary conditions on gauge fields in \eqref{gauge}; now we require
\begin{equation}
a_1=\mu_1+\frac{q_1}{r}+\calo(r^{-2})\,,\qquad \bar{a}_0=\bar{\mu}_0+\calo(r^{-2})\,,
\eqlabel{gaugemic}
\end{equation}
where the ``chemical potential'' parameters $\{\mu_1,\bar{\mu}_0\}$ are adjusted to keep fixed $U(1)_R$ charge density (parameter 
$q_1$), as well as the vanishing charge for the remaining $U(1)$ (in model DG-A).   For numerical analysis we set\footnote{Because 
we are discussing states in the CFT, the precise choice is irrelevant in so far as we represent the data in dimensionless quantities.} 
$\r_1=-q_1=\frac 12$ --- such a choice will put the phase transition numerically close to the one in grand canonical 
ensemble \eqref{tc} with $\mu_1=1$. 

\begin{figure}[t]
\begin{center}
\psfrag{s}{{$s/\r_1$}}
\psfrag{e}{{$\cale/\r_1^{3/2}$}}
\includegraphics[width=4in]{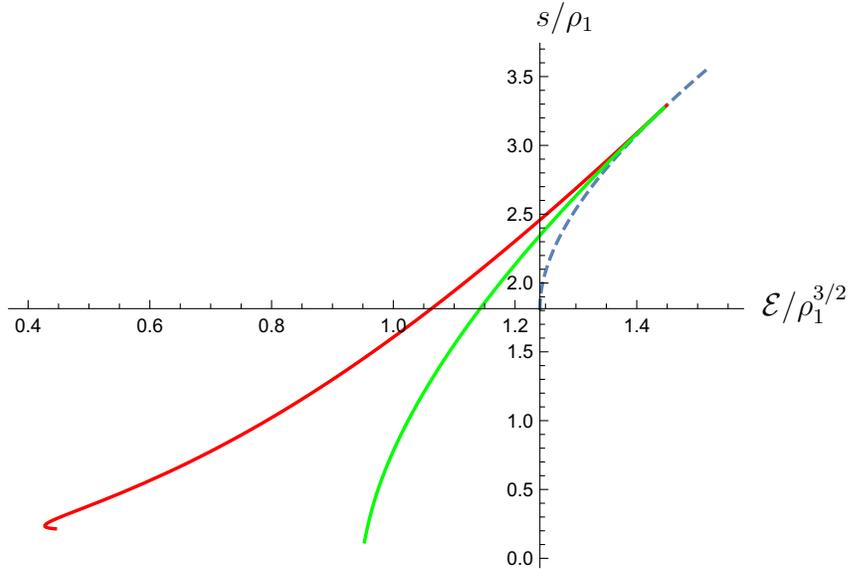}\qquad
\end{center}
  \caption{Entropy density $s$ as a function of the  energy density $\cale$ in model DG-A (green curve), DG-B (red curve) and the 
RN black brane (dashed blue curve).  } \label{figure2}
\end{figure}

Phase diagram in microcanonical ensemble is presented in fig.~\ref{figure2}. 
The dashed blue curve represents symmetry unbroken phase --- the RN black brane. 
The latter curve ends at the extremal RN solution:
\begin{equation}
\left\{\frac{\cale}{\r_{1}^{3/2}}\,, \frac{s}{\r_1}\right\}\bigg|_{RN,extremal} = \left\{\frac{2^{3/2}}{3^{3/4}}\,,\frac{2\pi}{3^{1/4}}
\right\}\,.
\eqlabel{extremal}
\end{equation}
In agreement with \eqref{tc}, the symmetry unbroken phase becomes unstable at 
\begin{equation}
\frac{\cale}{\r_1^{3/2}}\bigg|_{crit} = 1.448503(6)\,.
\eqlabel{ecrit}
\end{equation}
Two symmetry broken phases, \ie model DG-A (green curve) and model DG-B (red curve) dominate entropically 
the symmetry unbroken phase for $\cale< \cale_{crit}$. Interestingly, it is the exotic phase of the grand canonical ensemble 
(model DG-B) that is realized in the microcanonical description of the CFT below the critical energy density.

\subsection{Dynamics of DG-B model}

In the previous section we argued that DG-B model describes the dominant phase of the CFT below the critical energy density in 
microcanonical ensemble. Here we describe dynamics of spatially homogeneous and isotropic states of the model. 
We confirm that the criticality in the model is a standard holographic realization of the spontaneous symmetry breaking.

We begin with undoing the implicit bulk gauge symmetry fixing in the effective action of DG-B model \eqref{dglb}:
\begin{equation}
\begin{split}
&\call_{DG-B}=\frac 12 R -\left(\del \phi\right)^2-\sinh^2\phi \left(\del\theta -A^1\right)^2
-\frac14\ 
\left({F}^1\right)_{\mu\nu}\left({F}^1\right)^{\mu\nu}-V\,.
\end{split}
\eqlabel{dglb1}
\end{equation}
We further introduce a pair of scalar fields $f_i$ in lieu of $\{\phi,\theta\}$:
\begin{equation}
f_2+i f_1\equiv e^{i \theta}\ \sinh\phi \,.
\eqlabel{deffi}
\end{equation}
Assuming translational invariance along the spatial directions, we take 
\begin{equation}
\begin{split}
&ds_4^2=2dt\ \left(dr -A(t,r)\ dt\right)+\Sigma(t,r)^2\ \left[dx_1^2+dx_2^2\right]\,,\\
&A^1=a_1(t,r)\ dt\,,\qquad f_i=f_i(t,r) \,,
\end{split}
\eqlabel{anzatadgb}
\end{equation}
leading for the following equations of motion
\begin{equation}
\begin{split}
&0 =\Sigma''+\frac{\Sigma}{1+f_1^2+f_2^2}\left( (f_1 f_2'-f_2 f_1')^2+(f_1')^2+(f_2')^2\right) 
\,,\\
&0=a_1''+2 a_1' (\ln\Sigma)' -2 f_1' f_2 +2 f_1  f_2'\,,\\
&0=d_{+}'f_1 +\frac{d_{+}f_1}{1+f_1^2+f_2^2}  \left(
(\ln \Sigma)' -f_1  (f_1' (1+f_2^2)-f_2' f_1  f_2 )\right)
\\
&+\frac{d_{+}f_2  f_1}{1+f_1^2+f_2^2}  \left( (f_1' f_1  f_2 -f_2' (1+f_1^2)\right)
+\frac{f_1'}{\Sigma}  d_{+}\Sigma +a_1  \left(f_2  f_1  f_1'-(1+f_1^2) f_2'-f_2  (\ln\Sigma)' 
\right)\\
&-\frac12 f_2  a_1'+2f_1 (1+f_1^2+f_2^2) \,,\\
&0=d_{+}f_2'+\frac{d_{+}f_2}{1+f_1^2+f_2^2}  \left((\ln\Sigma)' +f_2  (f_1' f_1  f_2 -f_2' (1+f_1^2))\right)\\
&-\frac{d_{+}f_1 f_2}{1+f_1^2+f_2^2}  (f_1' (1+f_2^2)-f_2' f_1  f_2 )  
+\frac{f_2'}{\Sigma} d_{+}\Sigma 
+a_1  \left(f_1' (1+f_2^2)-f_2' f_1  f_2 +f_1  (\ln\Sigma)'\right)\\
& +\frac12 f_1  a_1'+2f_2 (1+f_1^2+f_2^2) \,,\\
&0=A''-2\frac{d_+\Sigma}{\Sigma^2}\ \Sigma'-(a_1')^2+\frac{2 d_{+}f_1}{1+f_1^2+f_2^2}  (f_1' (1+f_2^2)-f_2' f_1  f_2 )
\\&-\frac{2 d_{+}f_2}{1+f_1^2+f_2^2}  (f_1' f_1  f_2 -f_2' (1+f_1^2))-2a_1 (f_1' f_2 -f_1  f_2') \,,
\end{split}
\eqlabel{evolveoms}
\end{equation}
together with the constraint equations:
\begin{equation}
\begin{split}
&0=d_+'\Sigma+d_+\Sigma\ \left(\ln\Sigma\right)'-3 \Sigma+\Sigma\left(\frac 14 (a_1')^2-2f_1^2-2f_2^2\right) \;,\\
&0 =d_+^2\Sigma-2 A d_+'\Sigma-\frac{d_+\Sigma}{\Sigma^2}\
\left(A\Sigma^2\right)'+\frac{\Sigma}{1+f_1^2+f_2^2} \left(d_{+}f_1^2+d_{+}f_2^2+(f_1  d_{+}f_2 -f_2  d_{+}f_1 )^2\right)
\\&-\frac12 A  \Sigma  (a_1')^2+2 a_1  (d_{+}f_2  f_1 -f_2  d_{+}f_1 ) \Sigma +(f_1^2+f_2^2) \Sigma  a_1^2+2 (2 f_1^2+2 f_2^2+3) A  \Sigma \,,\\
&0=d_{+}'a_1+\frac{2 a_1'}{\Sigma} d_{+}\Sigma  -2 d_{+}f_2  f_1 +2 f_2  d_{+}f_1 -A' a_1'-2 a_1(f_1^2+f_2^2)  \,,
\end{split}
\eqlabel{coneoms2}
\end{equation}
where $'\equiv\del_r$ and $d_+\equiv \del_t+A\ \del_r$.
The constraint equations are preserved by the evolution equations provided they
are satisfied at a given timelike 
surface
--- which in our case is the AdS boundary.  

The general asymptotic boundary  ($r\to\infty$) solution  of the equations of motion, given by\footnote{Following
 \cite{Bosch:2016vcp}  
we used bulk gauge symmetry to require $a_1\sim \frac 1r$ as $r\to \infty$.}
\begin{equation}
\begin{split}
&\Sigma=r+\l(t)-\frac{f_{1,1}(t)^2+f_{2,1}(t)^2}{2r}+\calo(r^{-2})\,,\\
&A=\left(r+\l(t)\right)^2-f_{1,1}(t)^2-f_{2,1}(t)^2-\dot{\l}(t)+\frac{A_1(t)}{r}+\calo(r^{-2})\,,\\
&f_1=\frac{f_{1,1}(t)}{r}+\frac{f_{1,2}(t)}{r^2}+\calo(r^{-3})\,,\\
&f_2=\frac{f_{2,1}(t)}{r}+\frac{f_{2,2}(t)}{r^2}+\calo(r^{-3})\,,\\
&a_1=\frac{a_{1,1}(t)}{r}+\calo(r^{-2})\,,
\end{split}
\eqlabel{uvdgb}
\end{equation}
is characterized by seven (generically time-dependent) parameters
\begin{equation}
\{\l\,,\, f_{1,1}\,,\, f_{1,2}\,,\, f_{2,1}\,,\, f_{2,2}\,,\, A_{1}\,,\, a_{1,1}\}\,.
\eqlabel{uvdgbdyn}
\end{equation}  
The last two constraints in \eqref{coneoms2} imply that only five of them are independent:
\begin{equation}
\begin{split}
0=&\dot{A}_1+2 \dot{\lambda} (f_{1,1}^2+f_{2,1}^2)+2 \lambda (f_{1,1} \dot{f}_{1,1}+f_{2,1} \dot{f}_{2,1})
-f_{2,1} {\ddot f}_{2,1}-f_{1,1} \ddot{f}_{1,1} \\
&+2 f_{2,1} {\dot f}_{2,2}+2 f_{1,1} \dot{f}_{1,2}\,,\\
0=&\dot{a}_{1,1}-2 \dot{f}_{1,1} f_{2,1}+2 f_{1,1} \dot{f}_{2,1}-4 f_{1,1} f_{2,2}+4 f_{1,2} f_{2,1}\,.
\end{split}
\eqlabel{resconst}
\end{equation}
Restricting to static configurations, we identify $\{A_1,a_{1,1}\}$ with the energy density and the charge density as 
\begin{equation}
\cale=-2 A_1\,,\qquad \r_1=-a_{1,1}\,.
\eqlabel{iddyn}
\end{equation}
Dynamically, the constraint equations \eqref{resconst} become equivalent to energy and charge conservation, \ie
\begin{equation}
\dot{\cale}=0\,,\qquad \dot{\r}_1=0\,,
\eqlabel{dyncon}
\end{equation}
provided $f_{i,2}$ relate to $f_{i,1}$ as follows:
\begin{equation}
f_{i,2}(t)=\frac 12 \dot{f}_{i,1}(t)-\l(t) f_{i,1}(t)\,.
\eqlabel{fi2}
\end{equation}
As we will see, the boundary conditions \eqref{fi2} at late times,  $t\to \infty$, 
correctly reproduce the equilibrium thermodynamics of DG-B model 
discussed in section \ref{dgmicro}. Following the field redefinition \eqref{deffi} and 
relation to the expectation value of the dual operator \eqref{vevs},
we identify
\begin{equation}
\langle\calo_\phi\rangle   =  \sqrt{f_{1,1}(t)^2+f_{2,1}(t)^2} \,.
\eqlabel{ophitime}
\end{equation}
Additionally, $\mu_1$ in \eqref{gaugemic} is identified with   
\begin{equation}
\mu_1=\lim_{r\to\infty}\dot{\theta} = \frac {d}{dt}\ \biggl(\arctan \frac{f_{1,1}(t)}{f_{2,1}(t)}\biggr)\,.
\eqlabel{idmu1}
\end{equation}
Finally,   $\l(t)$ is the residual radial coordinate diffeomorphisms parameter 
\begin{equation}
r\to r+\l(t) \;,
\eqlabel{resdiffeo}
\end{equation}
which can adjusted to keep the apparent horizon at a fixed location, which in our
case will be $r=1$:
\begin{equation}
\biggl(\del_t +A(t,r)\ \del_r\ \biggr) \Sigma(t,r)\ \equiv\
 d_+\Sigma(t,r)\bigg|_{r=1}=0 \;.
\eqlabel{ldata}
\end{equation}

To initialize evolution at $t=0$, we  provide the bulk scalar profiles,
\begin{equation}
f_i(t=0,r)=\calo\left(\frac{1}{r}\right)\,,
\eqlabel{initfi}
\end{equation} 
along with the constant values of $\{A_1,a_{1,1}\}$, specifying the dual $CFT_3$
energy and charge densities according to \eqref{iddyn}. 
Eqs.~\eqref{evolveoms} are  employed to evolve such data \eqref{initfi} in time.

Further details of the numerical implementation can be found in Appendix \ref{dgbnum}.

\subsubsection{QNMs and linearized dynamics of DG-B model}\label{lindyn}

We discuss here the spectrum of QNMs associated with the spontaneous $U(1)_R$ 
symmetry breaking and the linearized dynamics in DG-B model. 

A symmetry unbroken phase of the boundary CFT is a $AdS_4$ RN black brane, which in 
notations \eqref{anzatadgb}, \eqref{redef}, \eqref{fieldsor} takes form :
\begin{equation}
\sigma(t,x)=\lambda\,,\qquad a(t,x)=\frac{x(4 m(\lambda x+1)+q_1^2 x )}{4(\l x+1)^2}\,,
\eqlabel{rnbh}
\end{equation}  
where $\l$ is determined from the stationarity of the horizon \eqref{ldata}
\begin{equation}
0=4 \lambda^2 (\lambda^2+4 \lambda+6)+(4 m+16) \lambda+q_1^2+4 m+4\,.
\eqlabel{lstatic}
\end{equation}
Introducing 
\begin{equation}
\begin{split}
&f_1(t,x) = -f(x)\ e^{\w_I t}\ \sin(\w_R t+g(x))\,,\qquad f_2(t,x) = f(x)\ e^{\w_I t}\ \cos(\w_R t+g(x))\,, \\
&g'(x)\equiv dg(x)\,,\qquad (\ln f(x))'\equiv \frac 1x+ ldf(x) \,,
\end{split}
\eqlabel{dgbqnm}
\end{equation}
where $\w=\w_R+i\ \w_I$ is the symmetry breaking QNM frequency,
we linearized the equations for the scalars $f_{i}$ on the gravitational 
background \eqref{rnbh} to find 
\begin{equation}
\begin{split}
&0=ldf'+ldf^2-dg^2+(4 \l^2 x^2 (\l^2 x^2+4 \l x+6)+4 x (m x^3+4) \l+q_1^2 x^4+4 m x^3+4)^{-1}\\
&\times  (-4 \w_I (\l x+1) (ldf \l x+ldf+\l)-4 dg (\l x+1)^2 \w_R-4 q_1 x (\l x+1) dg\\
&+(16 \l^2 x (\l^2 x^2+3 \l x+3)+(16 m x^3+16) \l+4 x^2 (q_1^2 x+3 m)) ldf+8 \l^2 (\l x+1)^2\\
&+2 x (4 \l m x+q_1^2 x+2 m))\,,
\\
&0=dg'+2 ldf dg+(4 \l^2 x^2 (\l^2 x^2+4 \l x+6)+4 x (m x^3+4) \l+q_1^2 x^4+4 m x^3+4)^{-1}\\
&\times  ((4 \w_R (\l x+1)) (ldf \l x+ldf+\l)-4 \w_I dg (\l x+1)^2+(16 \l^2 x (\l^2 x^2+3 \l x+3)\\
&+(16 m x^3+16) \l+4 x^2 (q_1^2 x+3 m)) dg+4 q_1 x (\l x+1) ldf+2 q_1 (2 \l x+1))\,.
\end{split}
\eqlabel{dgbqnmeoms}
\end{equation}
Eqns.~\eqref{dgbqnmeoms} are solved subject to regularity at the location of the black brane horizon 
($x=1$), and the asymptotic $AdS_4$ boundary conditions ($x\to 0$) following from  \eqref{fi2}
\begin{equation}
ldf=\frac 12\w_I-\l+\calo(x)\,,\qquad dg=-\frac 12\w_R+\calo(x)\,.
\eqlabel{uvgbqnm}
\end{equation}

\begin{figure}[t]
\begin{center}
\psfrag{w}{{$\w_R/\r_1^{1/2}$}}
\psfrag{a}{{$\w_I/\r_1^{1/2}$}}
\psfrag{m}{{$\cale/\r_1^{3/2}$}}
\includegraphics[width=2.7in]{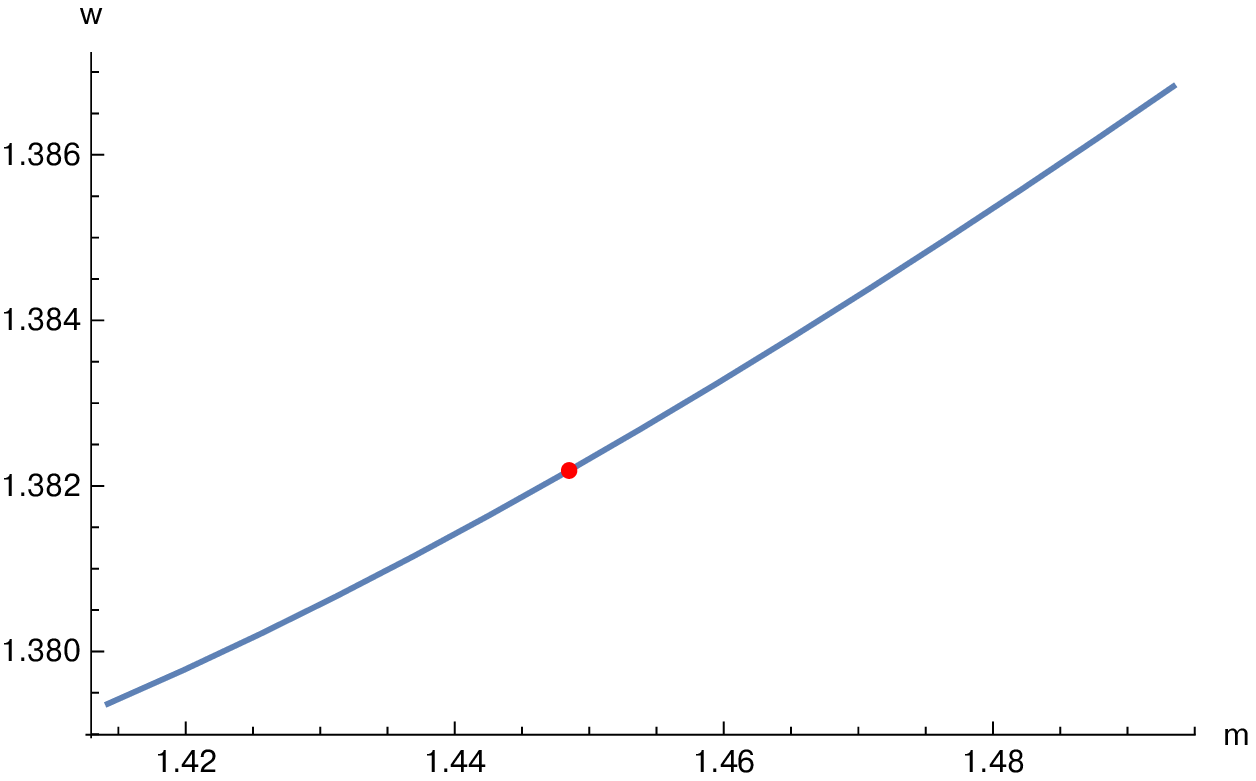}\qquad
\includegraphics[width=2.7in]{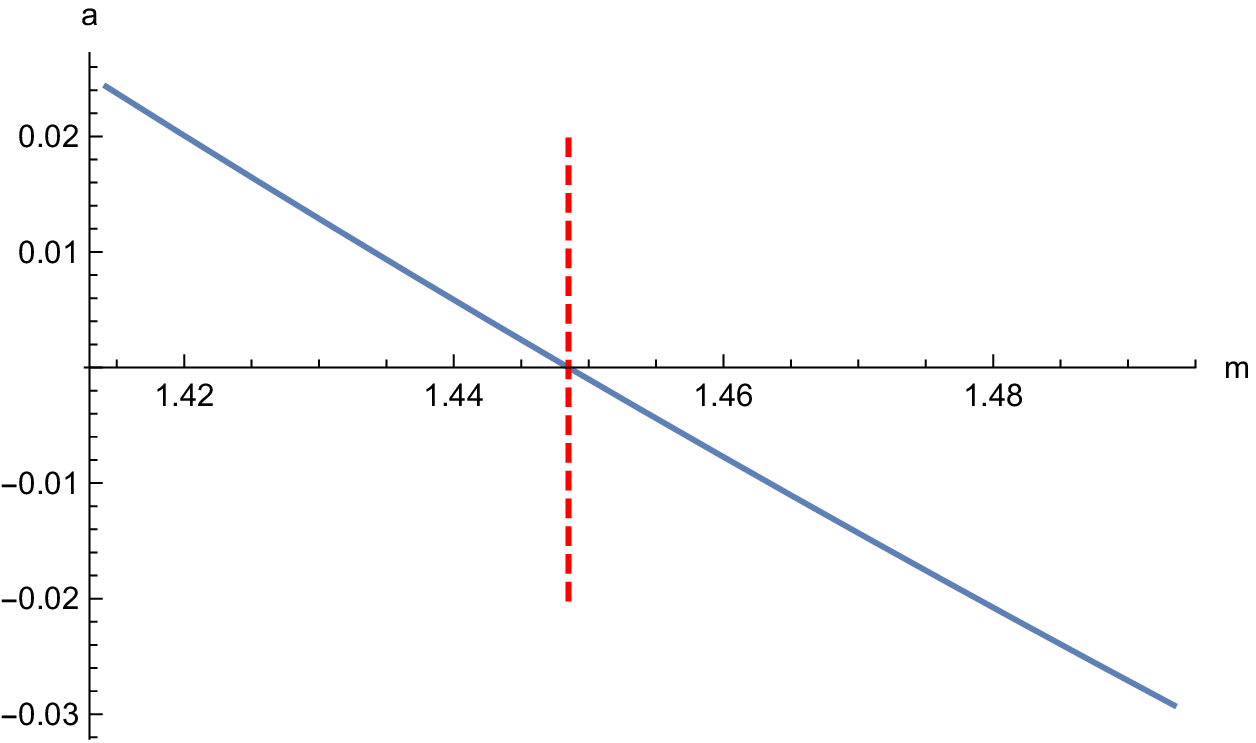}
\end{center}
  \caption{The spectrum of the $U(1)_R$ symmetry breaking fluctuations 
 at vanishing spatial momentum as a function of the energy density $\cale$. The red dot (left panel) 
represents $\mu_1$ at critical energy density, see \eqref{tc};
 the vertical red dashed line (right panel) represents the 
critical energy density, see \eqref{ecrit}.} \label{figure3}
\end{figure}

The spectrum of the symmetry breaking QNM is presented in fig.~\ref{figure3}. 
Note that $\w_I>0$, signalling the instability,  once $\cale<\cale_{crit}$, see  \eqref{ecrit}.

\begin{figure}[t]
\begin{center}
\psfrag{t}{{$t\r_1^{1/2}$}}
\psfrag{f}{{$f_{1,1}$}}
\psfrag{s}{{$\sqrt{f_{1,1}^2+f_{2,1}^2}$}}
\includegraphics[width=2.7in]{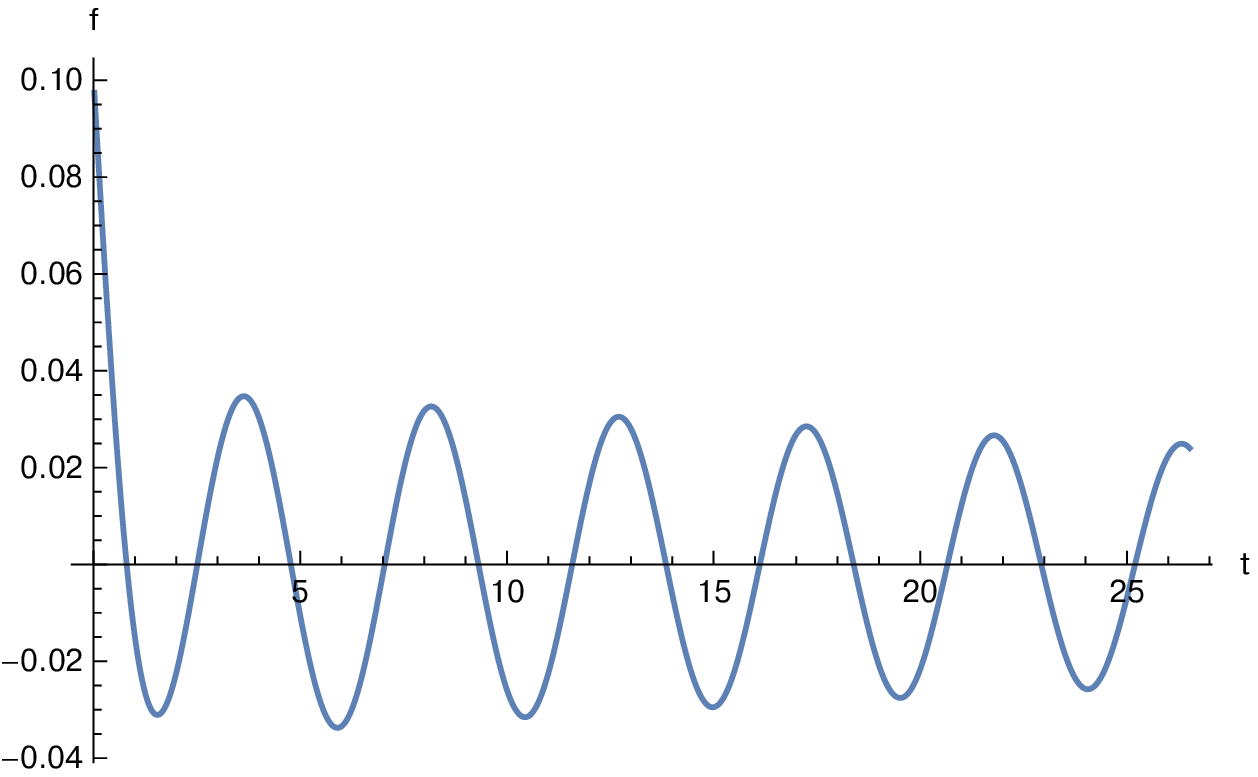}\qquad
\includegraphics[width=2.7in]{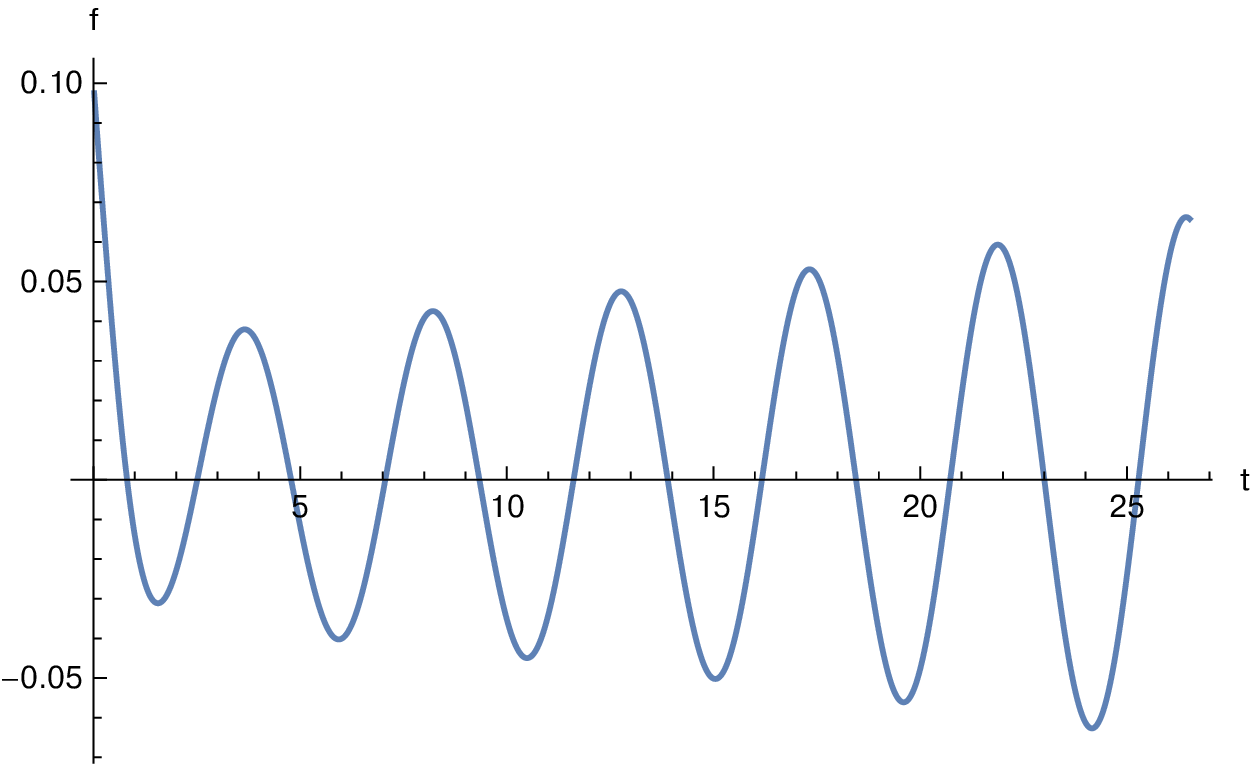}
\end{center}
  \caption{Linearized dynamics of the symmetry breaking scalars $f_{i}(t,x)$ on RN black brane background is captured 
by the evolution of  the expectation value $f_{1,1}(t)$, see \eqref{uvdgb}. Left panel 
corresponds to dynamics at energy density $\cale = 1.01538\cale_{crit}$  (symmetric phase is stable); 
right panel describes spontaneous symmetry breaking at   $\cale = 0.976328 \cale_{crit}$.  } \label{figure4}
\end{figure}

\begin{figure}[t]
\begin{center}
\psfrag{t}{{$t\r_1^{1/2}$}}
\psfrag{s}{{$\langle\calo_{\phi}\rangle$}}
\includegraphics[width=2.7in]{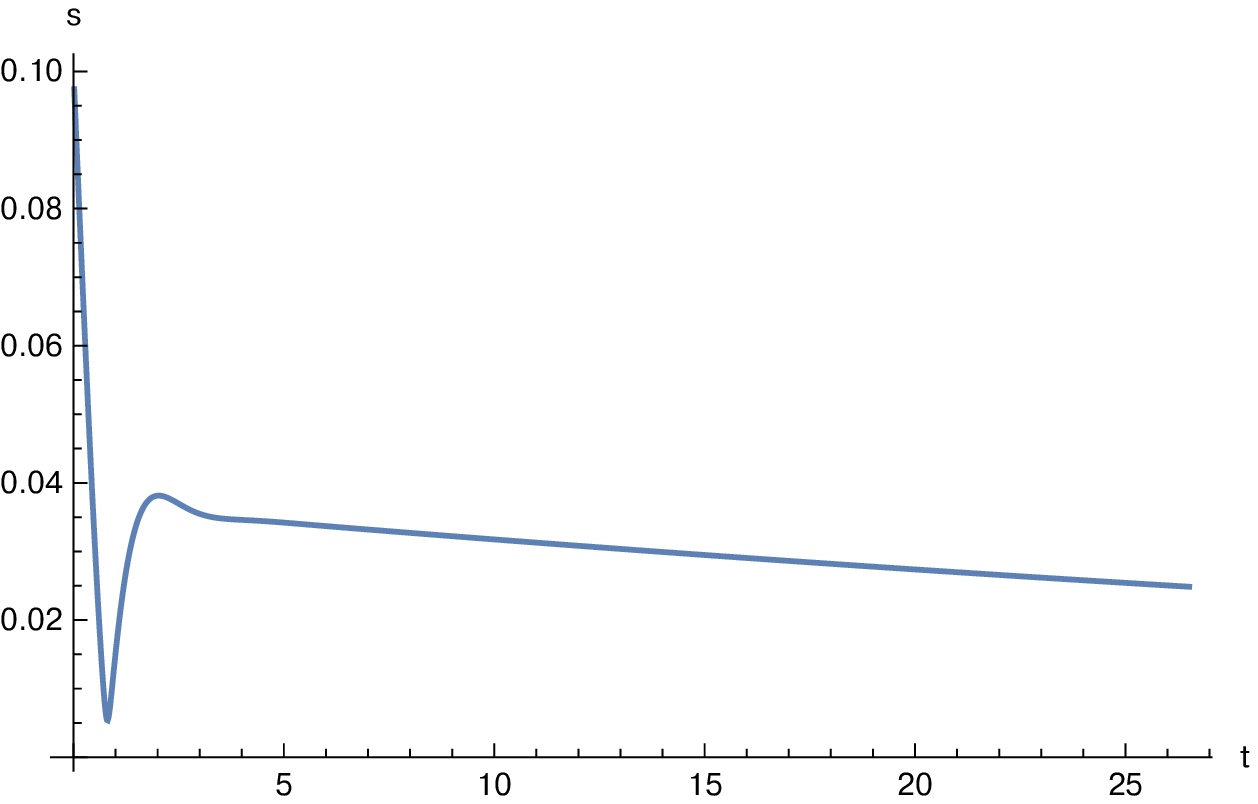}\qquad
\includegraphics[width=2.7in]{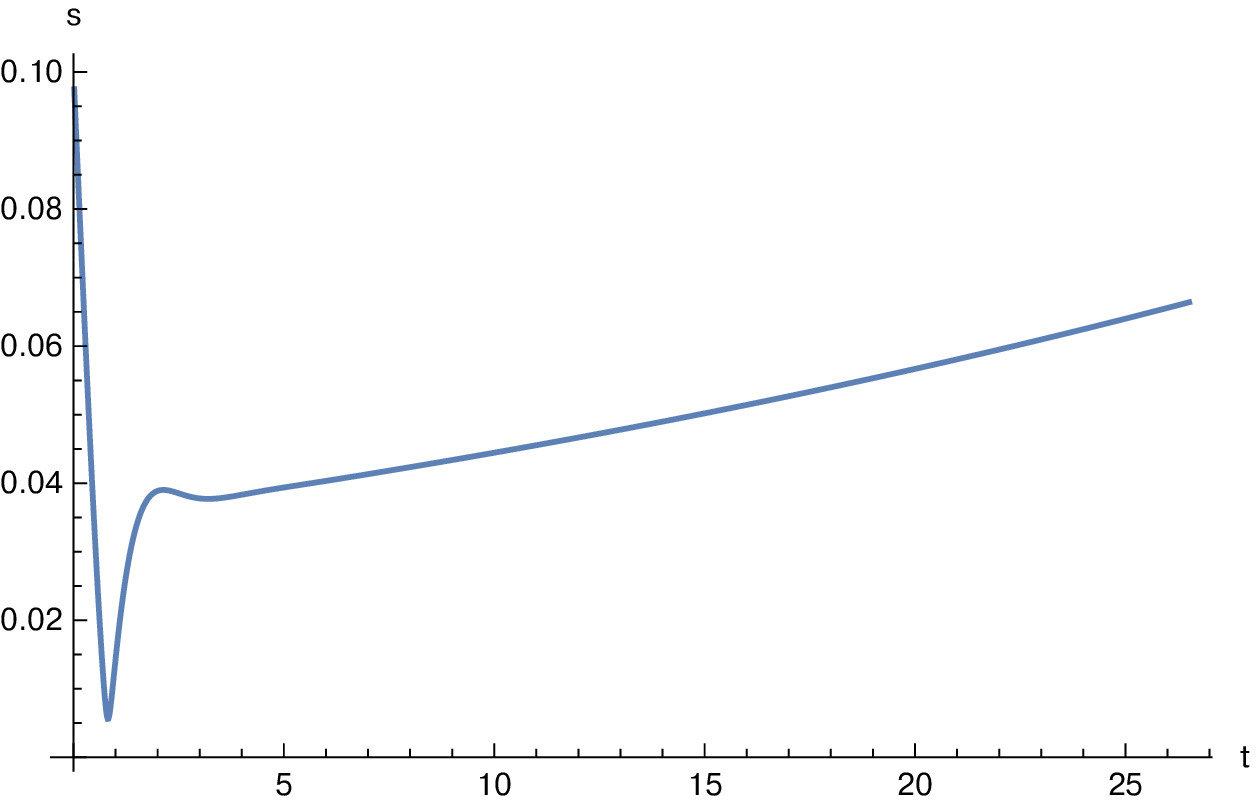}
\end{center}
  \caption{Same as in fig.~\ref{figure4}, except for the evolution of $\langle\calo_{\phi}\rangle=\sqrt{f_{1,1}^2+f_{2,1}^2}$.
The late time dynamics allows to extract $\w_{I}\equiv \w_{I,fit}$ to compare with the QNM behaviour,
see \eqref{dgbqnm}.} \label{figure5}
\end{figure}

As in \cite{Bosch:2017ccw}, we verify our dynamical code for DG-B by ``turning off'' 
the scalar fields backreaction on the geometry. Scalar evolution in this case must reproduce 
at late times the quasinormal behaviour \eqref{dgbqnm} with QNM spectrum presented in 
fig.~\ref{figure3}. Figs.~\ref{figure4}-\ref{figure5} present the evolution of   
$f_{1,1}(t)$ and $\langle\calo_{\phi}\rangle=\sqrt{f_{1,1}(t)^2+f_{2,1}(t)^2}$
for $\cale>\cale_{crit}$ (left panels --- stable case ) and $\cale<\cale_{crit}$ 
(right panels --- unstable case). From \eqref{dgbqnm} we expect 
\begin{equation}
\langle\calo_{\phi}\rangle\ \propto\ e^{\w_I t}\,.
\eqlabel{fitdgb1}
\end{equation}
Fitting the results in fig.~\ref{figure5} at late times and comparing with the 
expected decay/growth from the earlier QNM computations at the corresponding energy densities we find 
\begin{equation}
\bigg|\frac{\w_{I,fit}}{\w_{I,QNM}}-1\bigg|\lesssim 10^{-7}\,.
\eqlabel{fitdgb2}
\end{equation}

\subsubsection{Fully nonlinear evolution of DG-B model}

\begin{figure}[t]
\begin{center}
\psfrag{t}{{$t\r_1^{1/2}$}}
\psfrag{s}{{$\langle\calo_\phi\rangle$}}
\includegraphics[width=2.7in]{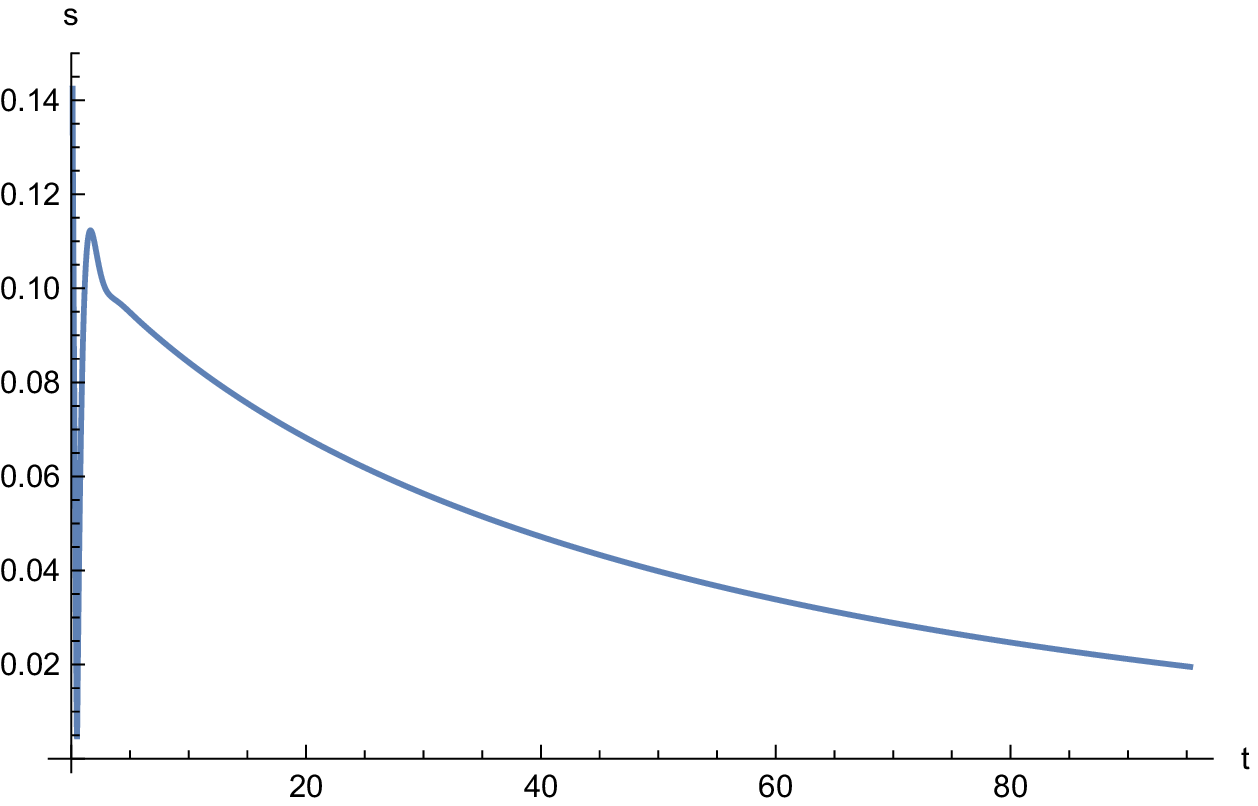}\qquad
\includegraphics[width=2.7in]{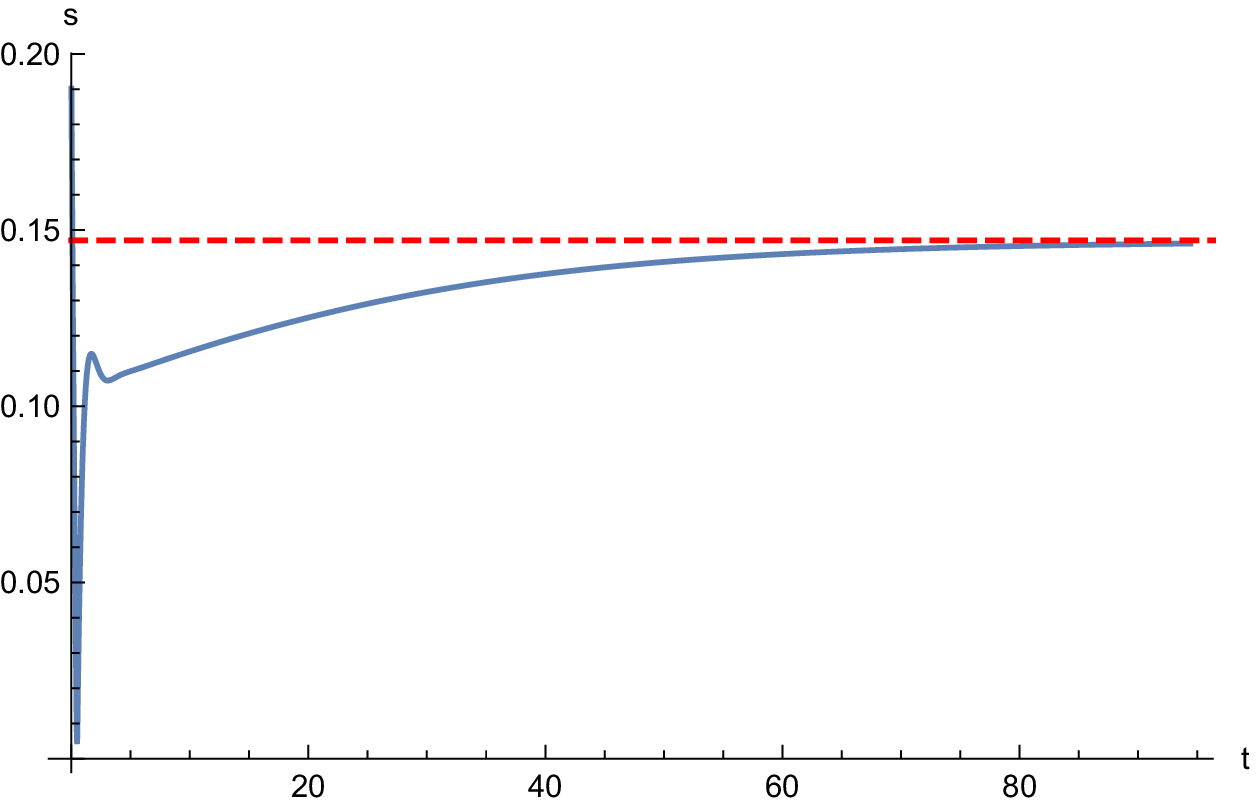}
\end{center}
  \caption{Fully nonlinear evolution of DG-B model. Left panel 
corresponds to dynamics at energy density $\cale = 1.01538\cale_{crit}$  (symmetric phase is stable); 
right panel describes spontaneous symmetry breaking at   $\cale = 0.976328 \cale_{crit}$.  The red dashed 
line represents the energy density corresponding asymptotic (equilibrium) expectation 
value of the symmetry breaking operator $\calo_\phi$
determined from microcanonical analysis of the model in section \ref{dgmicro}.} \label{figure6}
\end{figure}

Microcanonical analysis of the DG-B model in section \ref{dgmicro}
and the linearized dynamics discussed in section \ref{lindyn}
indicate that $U(1)_R$ symmetry is spontaneously broken at $\cale<\cale_{crit}$
in the model, while at $\cale>\cale_{crit}$ the symmetric phase is the dominant one. 
As fig.~\ref{figure6} presents, we find that this is indeed the case. Notice 
that approach to equilibrium is rather slow ($t_{equilibration} T \gtrsim 20 $ in this case ) --- as emphasized in \cite{Buchel:2015ofa} 
this is expected in the sector of the gauge theory responsible for the critical behaviour  close to transition.

We confirm the conclusion reached in section  \ref{dgmicro}: 
even though DG-B model exhibits an exotic thermodynamics discovered in 
\cite{Buchel:2009ge} in grand canonical ensemble, it represents the well-known 
holographic realization of the spontaneous symmetry breaking 
\cite{Hartnoll:2008vx,Hartnoll:2008kx} in microcanonical ensemble.

\section{Supersymmetric extension of BBL model}\label{bblsusy}

We would like to generalize BBL construction \cite{Bosch:2017ccw,Buchel:2009ge,Buchel:2010wk} to a ``supersymmetric'' model. 
As we show, this is rather easy to achieve. 

In four-dimensional gauged supergravity \cite{Ahn:2000mf,Bobev:2009ms} the Lagrangian of the scalar fields coupled to gravity is restricted to 
\begin{equation}
\call_{bosonic}=\frac 12 R +\call_{kin} -\calp\,,
\eqlabel{lag1}
\end{equation}
where the kinetic term is 
\begin{equation}
\call_{kin}=-K_{ij}\{\Phi_k\}\ \del_\mu\Phi^i \del^\mu \Phi^j\,,
\eqlabel{lag2}
\end{equation}
and the potential is determined from the (real) superpotential $\calw$
\begin{equation}
\calp= \left[K^{-1}\right]^{ij} \frac{\del \calw}{\del\Phi^i}\frac{\del \calw}{\del\Phi^j}-3 \calw^2\,.
\eqlabel{lag3}
\end{equation}
Assuming the metric ansatz,
\begin{equation}
ds_4^2=dr^2 +e^{2 A(r)}\left(-dt^2+dx_1^2+dx_2^2\right)\,,
\eqlabel{lag4}
\end{equation}
supersymmetric RG flow equations are obtained from the supersymmetry variation of a the fermions:
\begin{equation}
\frac{d\Phi^i}{dr}=\pm \left[K^{-1}\right]^{ij}\ \frac{\del \calw}{\del\Phi^j}\,,\qquad \frac{dA}{dr}=\mp \calw\,.
\eqlabel{lag5}
\end{equation}
Two comments are in order:
\begin{itemize}
\item the first order RG flow equations are consistent with the second order EOMs derived from the Lagrangian;
\item because flow equations for scalars are of the first order, supersymmetry imposes specific quantization for the scaling dimensions
for the dual operators (\eg see \eqref{dima} for DG-A model).
\end{itemize}

Recall, for the BBL model we have: 
\begin{equation}
\begin{split}
&\call_{kin}=-\frac 14 \left(\del\phi\right)^2-\frac 14 \left(\del\chi\right)^2\,,\qquad 
\calp_{BBL}= -3 -\frac 12\phi^2 +\chi^2 +\frac 12 g\ \phi^2 \chi^2 \,.
\end{split}
\eqlabel{bbl1}
\end{equation}
Note that the potential is unbounded from below for the nonlinear coupling $g<0$. The latter is necessary for the exotic phase structure 
of \cite{Buchel:2009ge,Buchel:2010wk}. For a supersymmetric generalization of the model, we introduce the superpotential 
$\calw$,
\begin{equation}
\calw=1+\frac 14 \phi^2 +\frac 12 \chi^2 +\frac{3}{160}\phi^4 
+\frac{3}{104}\chi^4+\left(\frac{1}{24}+\frac{g}{36}\right)\phi^2\chi^2\,,
\eqlabel{sbbl1}
\end{equation}
leading to 
\begin{equation}
\calp_{sBBL}=\calp_{BBL}+\calo(\phi^n \chi^{6-n})\,,\qquad n\ge 0\,,
\eqlabel{sbbl2}
\end{equation}
\ie the sBBL scalar potential captures the leading nonlinearity of the BBL model, but differs for higher order 
nonlinear interactions. As in BBL model, there is a single nonlinear coupling constant $g$. 
Explicitly, 
\begin{equation}
\begin{split}
&\calp_{sBBL}=\calp_{BBL}
-\frac{1}{116812800} \left(
520g \chi^2  \phi^2+540 \chi^4+780 \chi^2 \phi^2+351 \phi^4\right)^2
-\frac{9}{1600} \phi^6\\
&-\frac{45}{1352} \chi^6
-\frac{1}{12960} \chi^2 \phi^4 (-160 g^2-372 g+531)
 -\frac{1}{16848} \phi^2 \chi^4 (-208 g^2-84 g+1071)\,.
\end{split}
\eqlabel{sbbl3}
\end{equation}
Notice that additional terms in $\calp_{sBBL}$ (compare to $\calp_{BBL}$) 
are trivially unbounded from below 
when 
\begin{equation}
-\frac{21+27\sqrt{77}}{104}\ <\ g\ <\ 0\,.
\eqlabel{sbbl4}
\end{equation}

We explicitly verified that the second order equations of motion directly obtained from $\call_{sBBL}$ 
are consistent with the corresponding supersymmetric RG flow equations \eqref{lag5}. The supersymmetry 
imposes the following quantization on $\calo_r$, dual to the bulk scalar $\phi$ 
\begin{equation}
\Delta(\calo_r)=1\,.
\eqlabel{sbb5}
\end{equation}
Notice that the quantization \eqref{sbb5} is different from the one used in \cite{Bosch:2017ccw};
as a result, numerical implementation of sBBL model (see appendix \ref{numsbbl}) is closer to that of DG-B model
rather than the original BBL model. 

\subsection{Phase diagram and QNMs}\label{msbbl}

\begin{figure}[t]
\begin{center}
\psfrag{e}{{$\frac{2\k^2 \cale}{\Lambda^3}$}}
\psfrag{s}{{$\k^2 s_{sym}/(2\pi \Lambda^2)$}}
\psfrag{g}{{$g$}}
\psfrag{c}{{$\frac{2\k^2 \cale_{crit}}{\Lambda^3}$}}
  \includegraphics[width=2.5in]{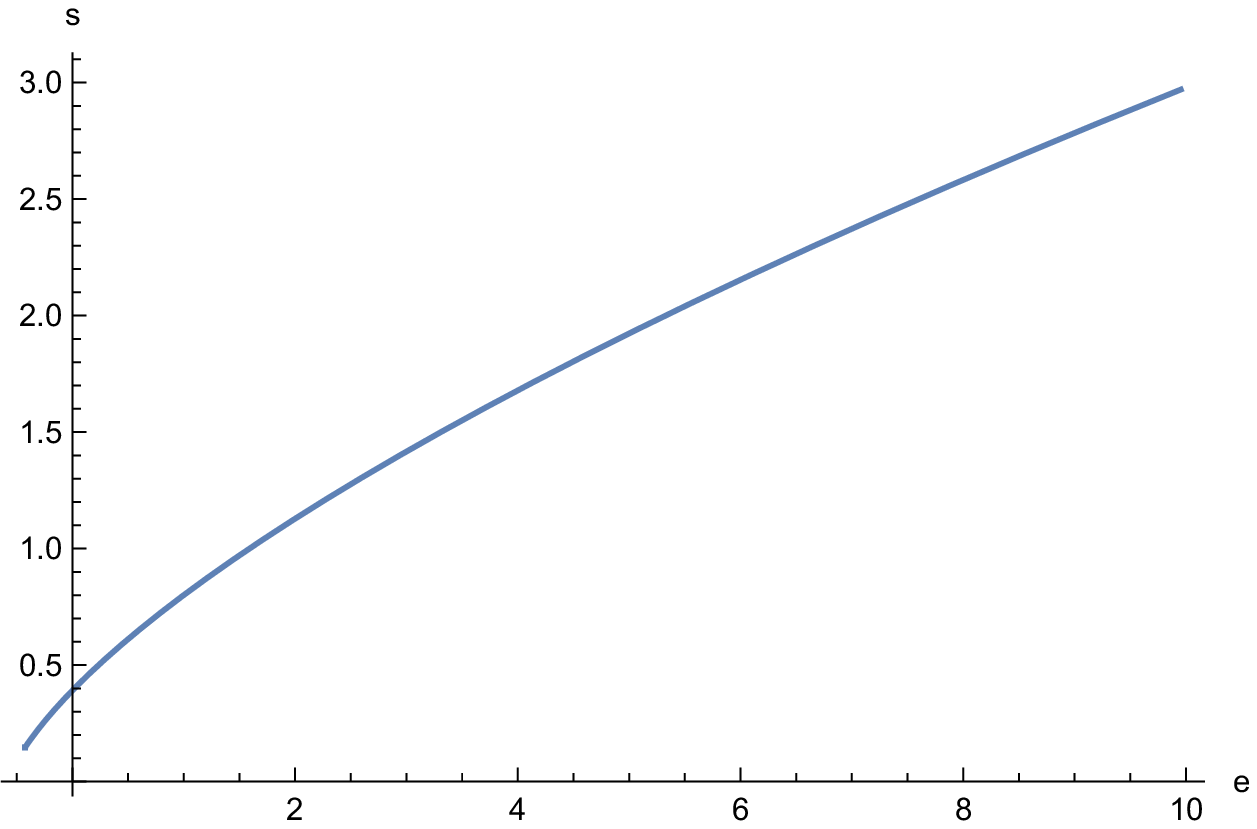}\,\,\,\,\,\,\,\,\,
  \includegraphics[width=2.5in]{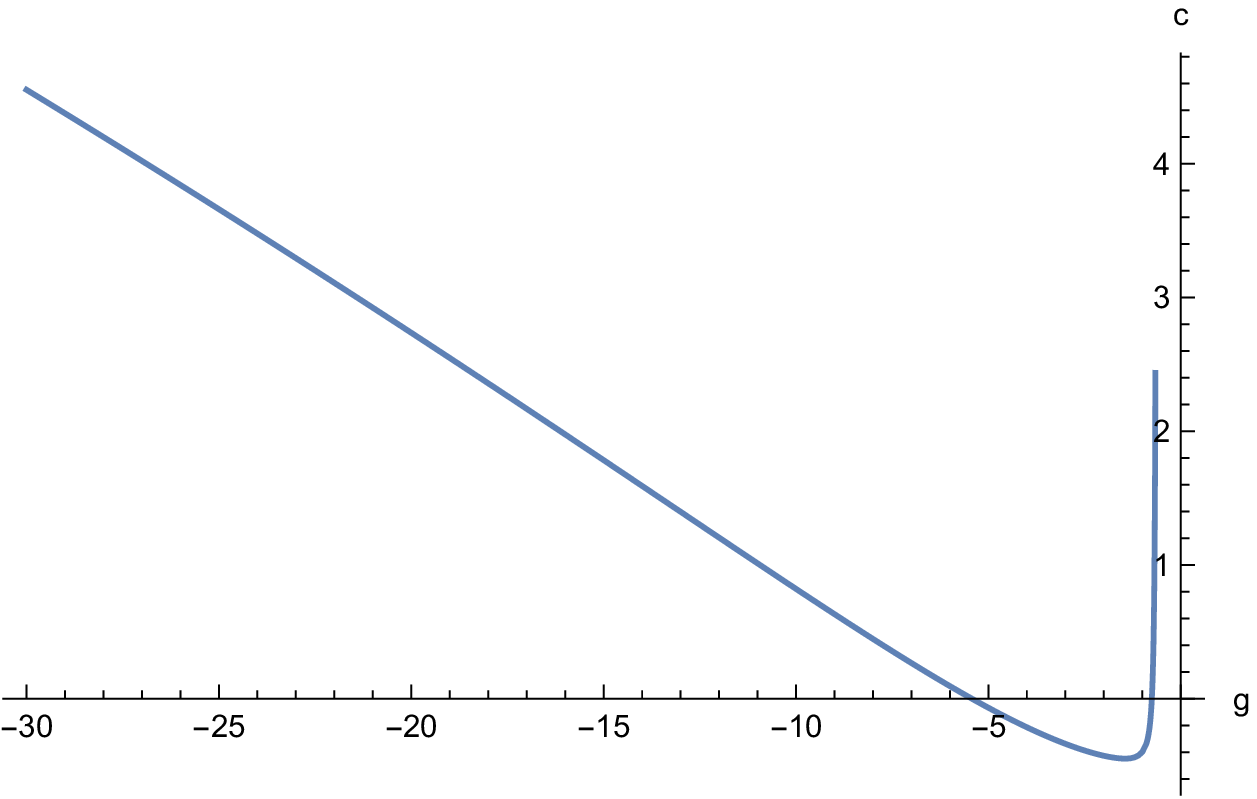}
\end{center}
 \caption{  Entropy density $s_{sym}$ of the $\zet_2$-symmetric phase, \ie with $\langle\calo_i\rangle=0$, 
of sBBL model  as a function of energy density $\cale$ (left panel).  As the energy density 
is decreased below the critical one $\cale_{crit}$, the symmetric phase becomes perturbatively unstable with
respect to linearized $\zet_2$-symmetry breaking fluctuations. The critical energy 
density for the onset of this instability as a function of the nonlinear coupling 
$g$ (see \eqref{sbbl3}) (right panel). } \label{figure7}
\end{figure}

\begin{figure}[t]
\begin{center}
\psfrag{e}{{$\frac{2\k^2 \cale}{\Lambda^3}$}}
\psfrag{w}{{$\frac{\Im{(\w)}}{\Lambda}$}}
\psfrag{r}{{$\frac{\Re{(\w)}}{\Lambda}$}}
  \includegraphics[width=2.5in]{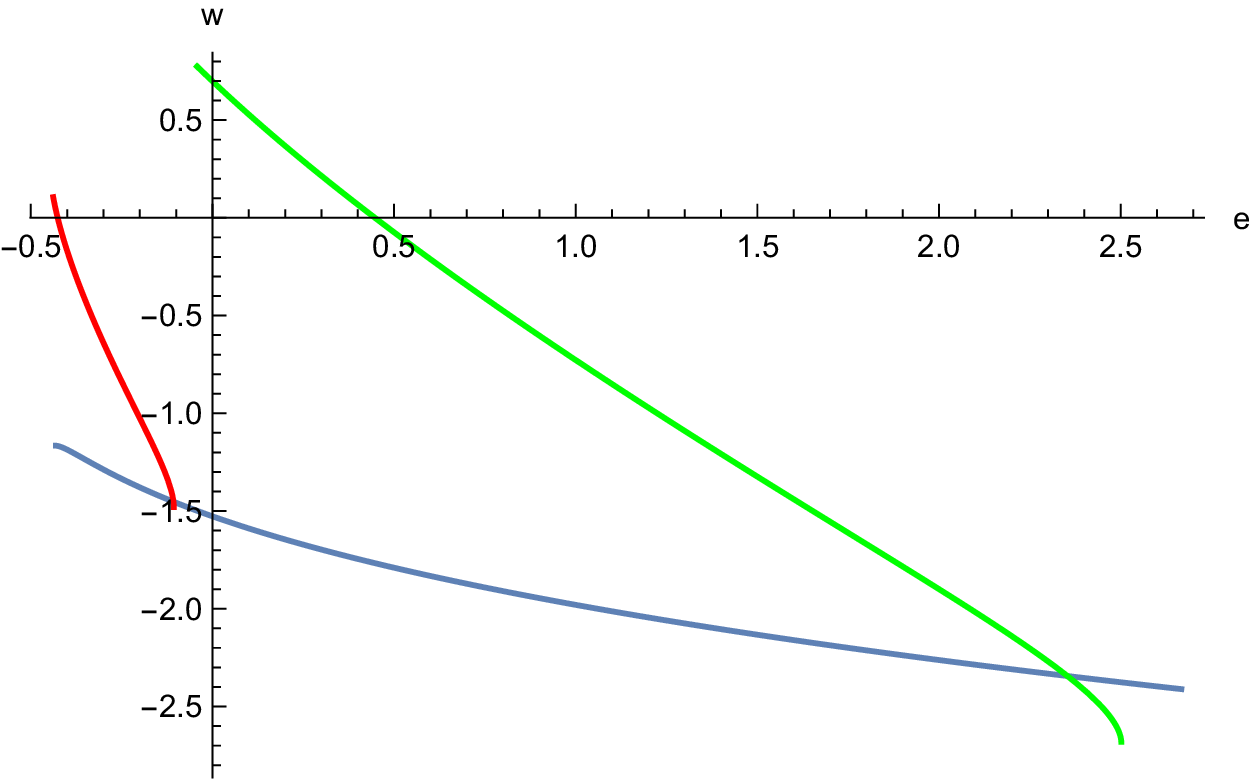}\,\,\,\,\,\,\,\,\,
  \includegraphics[width=2.5in]{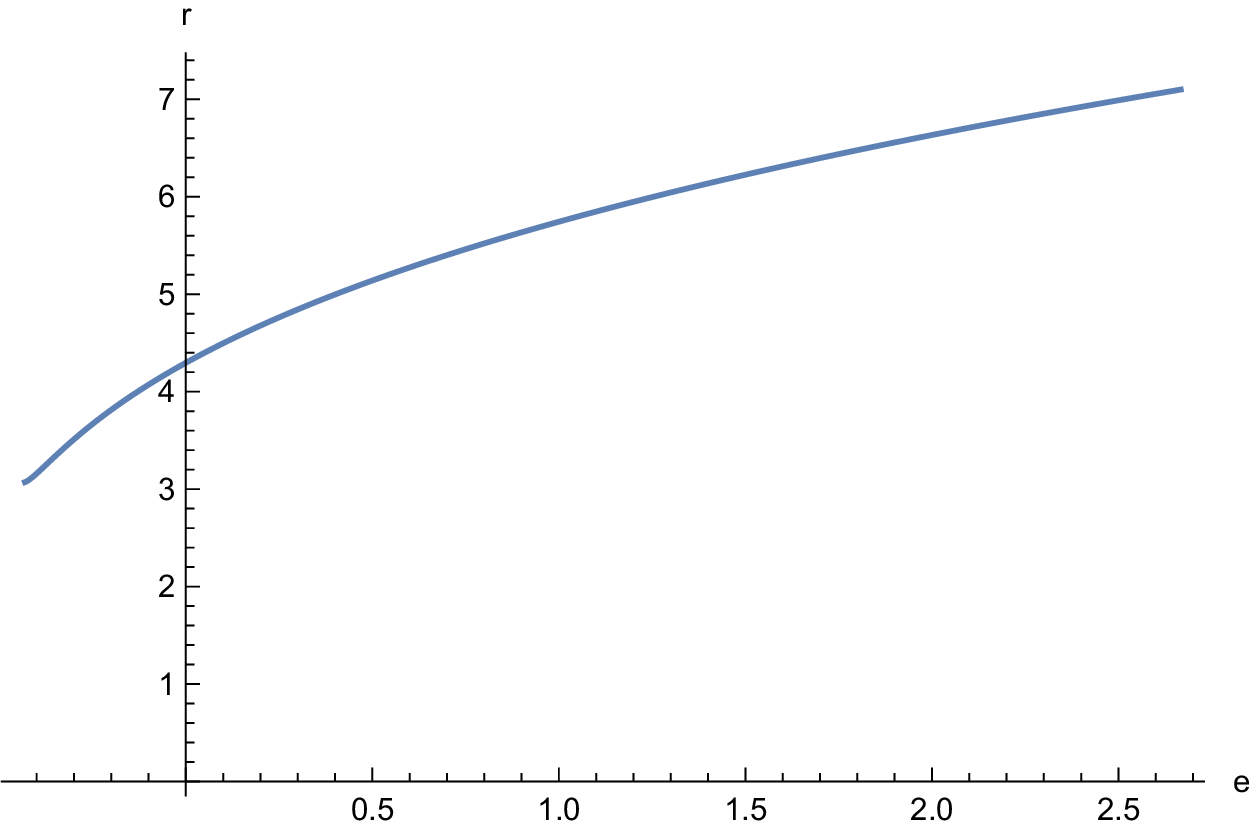}
\end{center}
 \caption{ Left panel: spectrum of $\zet_2$ symmetry breaking QNM in sBBL model at 
$g=-2$ (red curve) and $g=-8$ (green curve)  at vanishing spatial momentum. 
These QNMs have vanishing real part. The blue curves 
(both panels) represent the spectrum of QNM in sBBL model that connect to standard $\Delta=1$ QNM of $AdS_4$ 
black brane in the limit $\cale/\Lambda^3\to \infty$.} \label{figure8}
\end{figure}

\begin{figure}[t]
\begin{center}
\psfrag{e}{{$\frac{2\k^2 \cale}{\Lambda^3}$}}
\psfrag{s}{{$\k^2 s_{sym}/(2\pi\Lambda^2)$}}
  \includegraphics[width=4in]{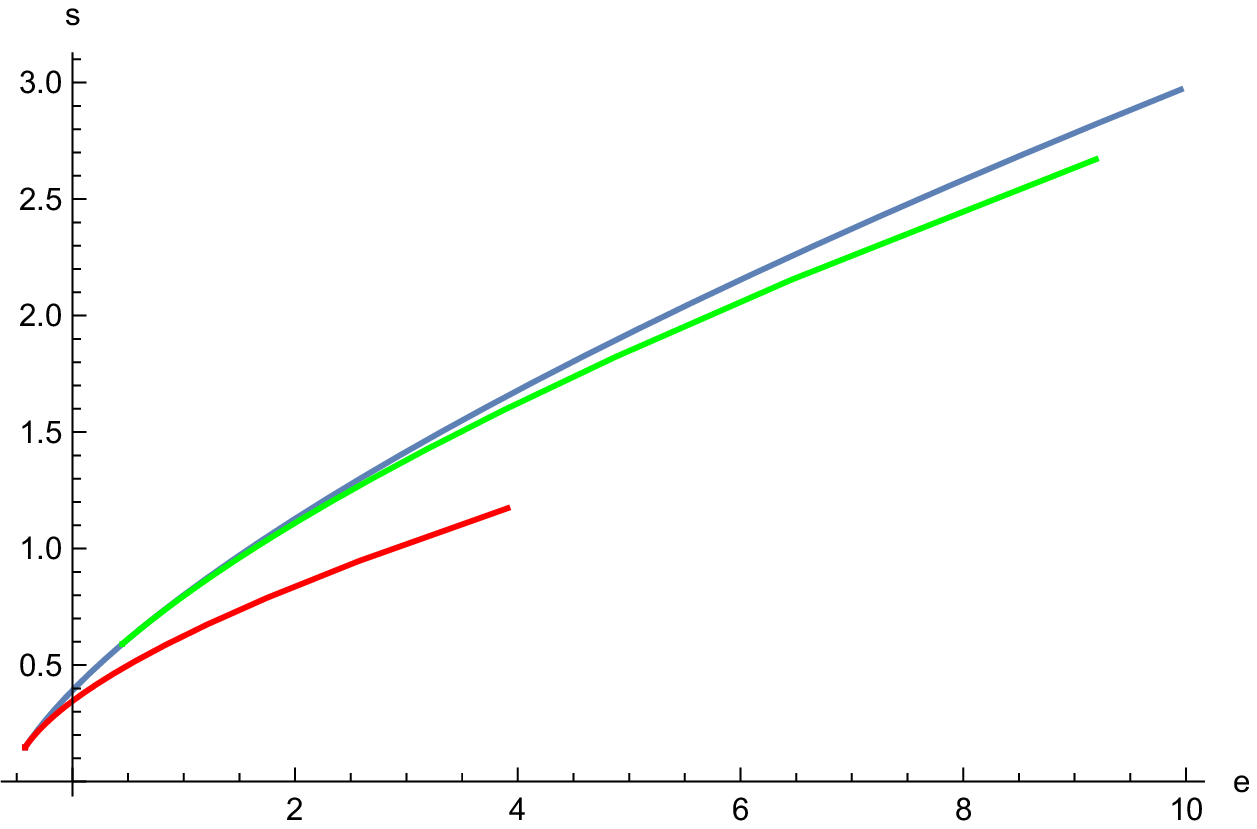}
\end{center}
 \caption{Phase diagram of sBBL model in microcanonical ensemble. The blue curve 
represents the symmetric phase of the system; the green and red curves are the exotic phases 
at $g=-8$ and $g=-2$ correspondingly, which exist only for $\cale>\cale_{crit}$.  } \label{figure9}
\end{figure}

Since BBL and sBBL models differ by higher-order nonlinearities in bulk scalar potentials,
holographic renormalization for the models is the same. In particular, we can borrow the 
expressions for the thermodynamic quantities in \cite{Buchel:2009ge}, appropriate for the supersymmetric 
quantization \eqref{sbb5}. We present the results only --- for detailed discussion follow 
\cite{Buchel:2009ge,Bosch:2017ccw,Buchel:2010wk}. 
\begin{itemize}
\item There are two equilibrium phases of the sBBL model, distinguished by the symmetry property 
under $\chi\leftrightarrow -\chi$: the symmetric phase with $\langle\calo_i\rangle=0$, and the symmetry broken phase 
with  $\langle\calo_i\rangle\ne 0$.
\item The entropy density of the symmetric phase $s_{sym}$ as a function of the energy density $\cale$ is presented in 
figure~\ref{figure7} (left panel). While this phase is thermodynamically stable 
$\frac{\del^2 \cale}{\del s_{sym}^2}>0$, it is 
perturbatively unstable with respect to a linearized symmetry breaking fluctuations. 
The critical energy for the instability as a function of the coupling constant $g$ in \eqref{sbbl3} 
is shown in the right panel. 

Notice that the energy density of the symmetric phase (as well as the critical energy density) can be 
negative. Typically, finite counterterms in holographic renormalization make the definition of the 
energy density ambiguous. This is the case for the $\phi$ scalar quantization with 
$\Delta(\calo_r)=2$: a finite boundary counterterm\footnote{$\sqrt{-h}$ is the induced metric on 
the cut-off surface $\del\calm_4$ in the holographic renormalization.} 
\begin{equation}
\call_{counter,finite}\propto \int_{\del\calm_4} dx^3 \sqrt{-h} \phi^3\,,
\eqlabel{finite}
\end{equation} 
renders the definition of the energy density $\cale$ ambiguous. However, for the supersymmetric 
quantization, \ie \eqref{sbb5}, such term is not allowed, as it will violate the energy conservation Ward
identity. So, the negative energy densities in the symmetric phase of the sBBL model 
are unambiguous\footnote{It is easy to verify that for the supersymmetric 
RG flows \eqref{lag5} in sBBL model the (vacuum) energy density vanishes, as required by the boundary CFT
supersymmetry. }. We discuss dynamics of sBBL model in section \ref{susybbldyn} for two representative values of 
the coupling constant $g=\{-2,-8\}$. Notice that $\cale_{crit}(g=-2)<0$ and $\cale_{crit}(g=-8)>0$. 
Additionally, the value $g=-2$ is within the range \eqref{sbbl4}, where the gravitational 
scalar potential $\calp_{sBBL}$
is trivially unbounded from below.
\item  The spectrum of $\zet_2$ symmetry breaking quasinormal modes in sBBL model 
is presented in fig.~\ref{figure8} as a function of energy density for  $g=-2$ (red curve) 
and $g=-8$ (green curve) at vanishing spatial momentum. In agreement with 
results in fig.~\ref{figure7} (right panel), these modes 
become unstable \ie have a positive imaginary part, when $\cale<\cale_{crit}$ with 
\begin{equation}
\frac{2\k^2 \cale_{crit}}{\Lambda^3}\bigg|_{g=-2}=-0.42618(8)\,,\qquad 
\frac{2\k^2 \cale_{crit}}{\Lambda^3}\bigg|_{g=-8}=0.44726(8)\,.
\eqlabel{ecrit28}
\end{equation}
The unusual feature of these modes, first observed in  \cite{Buchel:2010wk}, is the fact that they  
have vanishing real part, \ie
\begin{equation}
\Re{(\w)}\bigg|_{red\ \&\ green} = 0\,.
\eqlabel{noreal}
\end{equation}
Because of \eqref{noreal}, these modes must disappear 
from the spectrum in the limit $\frac{\cale}{\Lambda^3}\to \infty$, \ie when sBBL approaches its UV 
(conformal) fixed point\footnote{Recall that QNMs of a holographic dual to a CFT have both 
real and imaginary parts, typically with $\Re(\w)\sim -\Im (\w)$ \cite{Nunez:2003eq}.}. 
The blue curves in fig.~\ref{figure8} represent the sBBL QNM mode, which connects 
at asymptotically large energies to $\Delta=1$ QNM of the $CFT_3$. Notice that as 
$\cale$ increases sufficiently far over the corresponding $\cale_{crit}$, there is a 
"level crossing'' (left panel), and the red and green curve QNMs cease to dominate the 
relaxation of the symmetry breaking fluctuations in the system --- it is governed by the blue 
curve QNM. 
\item As in BBL model, in sBBL model new phases with spontaneous $\zet_2$ symmetry breaking 
appear in the microcanonical ensemble once $\cale> \cale_{crit}$,
fig.~\ref{figure9}. Once again, these new phases have lower entropy density than the symmetric phase 
(blue curve) and thus never dominate the dynamics of the system. 
\end{itemize}

\subsection{Dynamics of sBBL horizons}\label{susybbldyn}

In  section \ref{msbbl} 
we established that the equilibrium physics of BBL and sBBL models in microcanonical ensemble 
is identical: in both models the $\zet_2$ symmetric equilibrium phase becomes unstable below some 
critical energy density $\cale_{crit}$; there is no end-point for the instability and the 'hairy' phases bifurcate
from the onset of the instability towards $\cale>\cale_{crit}$. These new phases are exotic --- they have lower 
entropy density than the symmetry preserving phase and thus never dominate dynamically. 

We now study the dynamics of the sBBL model. We highlight the main results and refer the reader to 
\cite{Bosch:2017ccw} for further  implementation details. Because 
the quantization of the operator $\calo_r$  dual to the bulk scalar $\phi$ is modified, 
see \eqref{sbb5}, the general asymptotic boundary ($r\to \infty$) 
solution of the equations of motion is now given by
\begin{equation} 
\begin{split}
&\Sigma=r+\l(t)-\frac 18 p_1(t)^2\ \frac 1r +\calo\left(\frac{1}{r^2}\right)\,,\\
&A=\frac{r^2}{2}+\l(t)\ r -\frac 18 p_1(t)^2+\frac 12\l(t)^2
-\dot\l(t) + \frac {\mu}{r}
+\calo\left(\frac{1}{r^2}\right) \,,\\
&\phi=\frac{p_1(t)}{r}+\left(p_2+\dot{p}_1(t)-\l(t) p_1(t)\right)
\frac{1}{r^2}+\calo\left(\frac{1}{r^3}\right)\, , \\
&\chi=\frac{q_4(t)}{r^4}+\calo\left(\frac{1}{r^5}\right) \, .
\end{split}
\eqlabel{uvsbbl}
\end{equation}
It is characterized by two constants $\{p_2,\mu\}$, and three dynamical variables 
$\{p_1(t), q_4(t), \l(t)\}$. These parameters have the following interpretation:
\begin{itemize}
\item $p_2$ and $p_1(t)$ are identified with the deformation mass scale $\Lambda$ and the 
expectation value of the relevant operator $\calo_r$ of the dual $QFT_3$,
\begin{equation}
p_2=2^{-2/3}\ \Lambda^2\,,\qquad
p_1(t)=2^{-1/3}\ \langle\calo_r(t)\rangle\,;
\eqlabel{phidat2}
\end{equation}
\item $q_4(t)$ is the normalizable coefficient of the bulk scalar $\chi$, identified 
with the expectation value of the $\zet_2$-symmetry breaking irrelevant operator $\calo_i$
of the dual $QFT_3$,
\begin{equation}
q_4(t)=\langle\calo_i(t)\rangle\,; 
\eqlabel{chidat2}
\end{equation}
\item $\mu$ is related to the conserved energy density $\cale$ of the boundary $QFT_3$ as follows
\begin{equation}
\frac{2\k^2 \cale}{\Lambda^3}=\frac{-4\mu}{\Lambda^3} \;;
\eqlabel{endata2}
\end{equation}
\item   $\l(t)$ is the residual radial coordinate diffeomorphisms parameter 
\begin{equation}
r\to r+\l(t) \;,
\eqlabel{resdiffeo2}
\end{equation}
which can adjusted to keep the apparent horizon at a fixed location, which in our
case will be $r=1$:
\begin{equation}
\biggl(\del_t +A(t,r)\ \del_r\ \biggr) \Sigma(t,r)\ \equiv\
 d_+\Sigma(t,r)\bigg|_{r=1}=0 \;.
\eqlabel{ldata2}
\end{equation}
\end{itemize}

To initialize evolution at $t=0$, we  provide the bulk scalar profiles,
\begin{equation}
\phi(t=0,r)=\calo\left(\frac{1}{r}\right)\,,\qquad
\chi(t=0,r)=\calo\left(\frac{1}{r^4}\right) \;,
\eqlabel{initphichi}
\end{equation} 
along with the values of $\{p_2,\mu\}$, specifying the dual $QFT_3$ mass scale $\Lambda$ \eqref{phidat2} 
and the initial state energy density  
$\cale$ \eqref{endata2}.

Some details for the modification of the numerical code of \cite{Bosch:2017ccw}
are collected in appendix \ref{numsbbl}.

\subsubsection{Dynamics of symmetric sBBL sector and its linearized symmetry breaking fluctuations}\label{ldynsbbl}

\begin{figure}[t]
\begin{center}
\psfrag{t}{{$t\Lambda$}}
\psfrag{p}{{$\langle \calo_r\rangle/\Lambda$}}
\psfrag{l}{{$\ln|\langle \calo_r-\calo_r^e\rangle/\Lambda|$}}
  \includegraphics[width=2.5in]{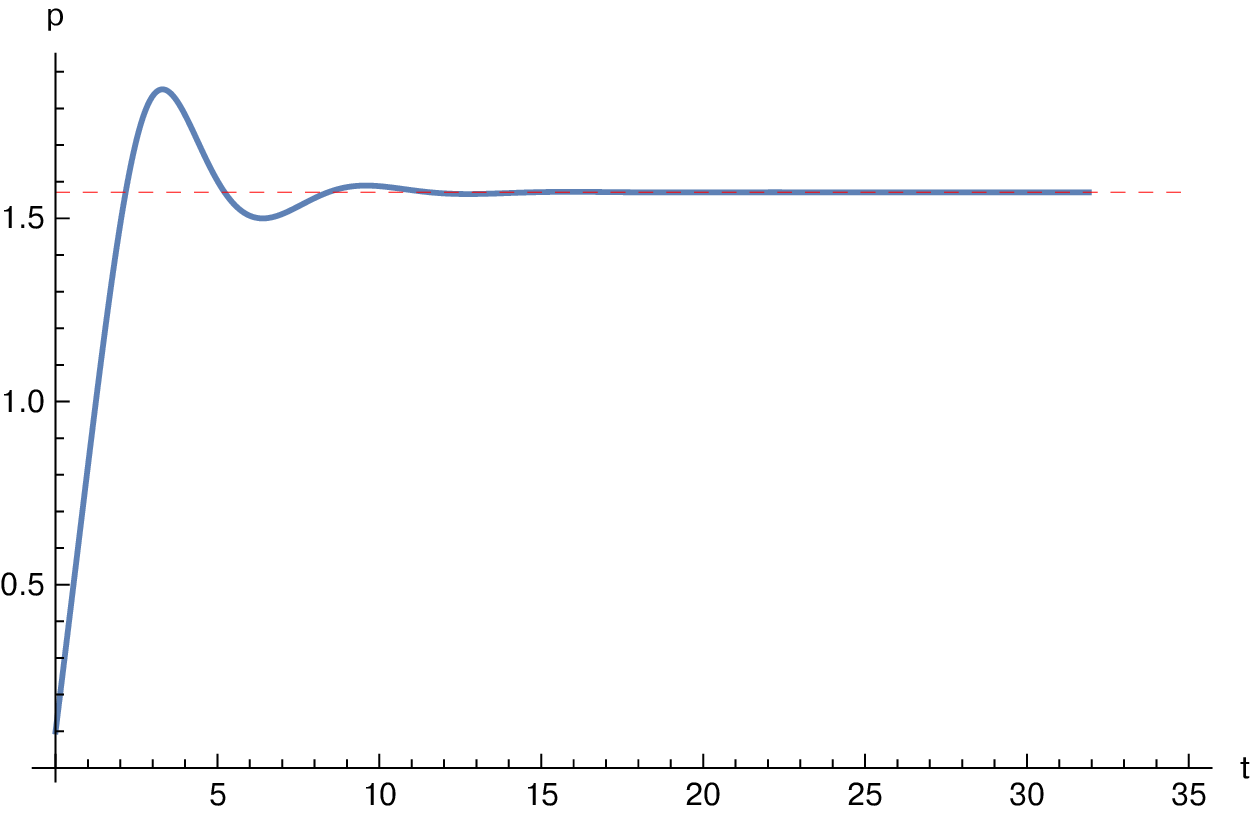}\,\,\,\,\,\,\,\,\,
  \includegraphics[width=2.5in]{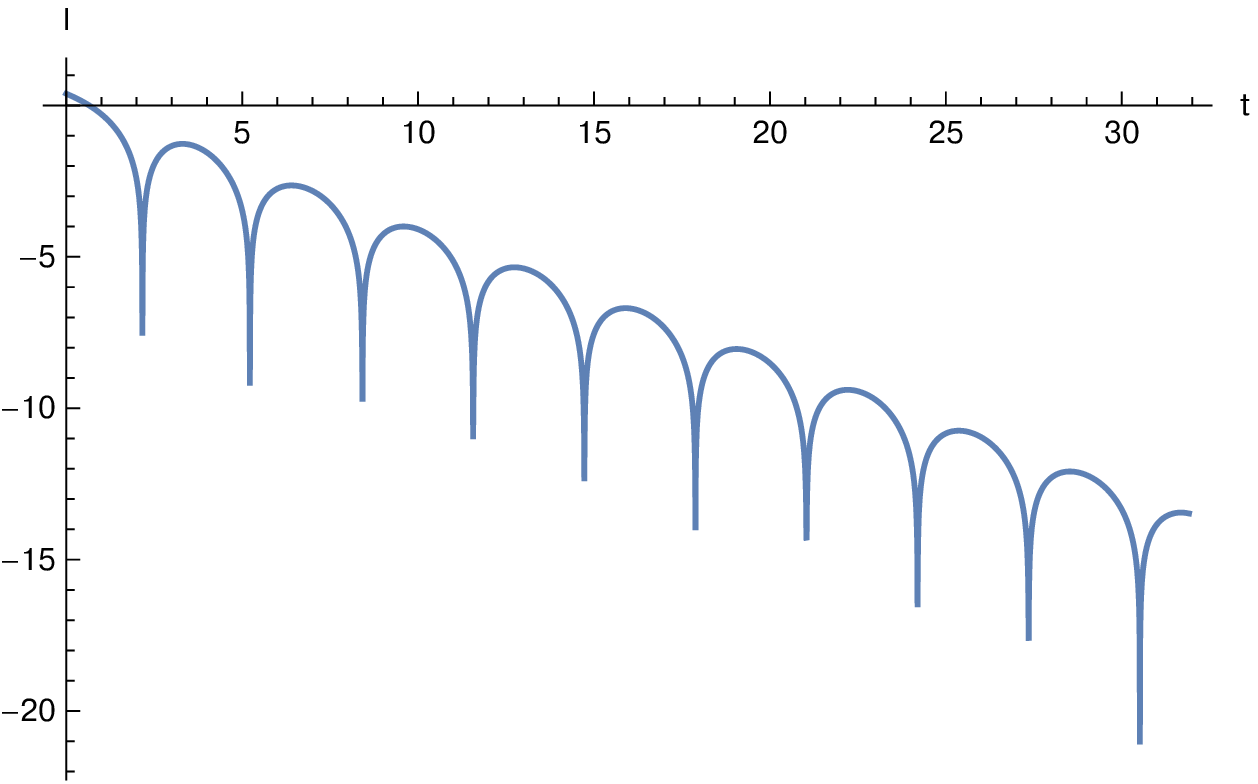}
\end{center}
 \caption{Typical relaxation in $\zet_2$ symmetric sector of sBBL model. The expectation 
value of $\calo_r$ operator (dual to the bulk scalar $\phi$)
settles to its equilibrium value $\calo_r^e$ denoted by a dashed red line (left panel).
Right panel shows a characteristic QNM ring-down of the expectation value to
its equilibrium value.  } \label{figure10}
\end{figure}

\begin{figure}[t]
\begin{center}
\psfrag{t}{{$t\Lambda$}}
\psfrag{a}{{$\k^2 s_{sym}/(2\pi\Lambda^2)$}}
\psfrag{d}{{$(\dd \dot{s}_{sym})/(\Lambda s_{sym})$}}
  \includegraphics[width=2.5in]{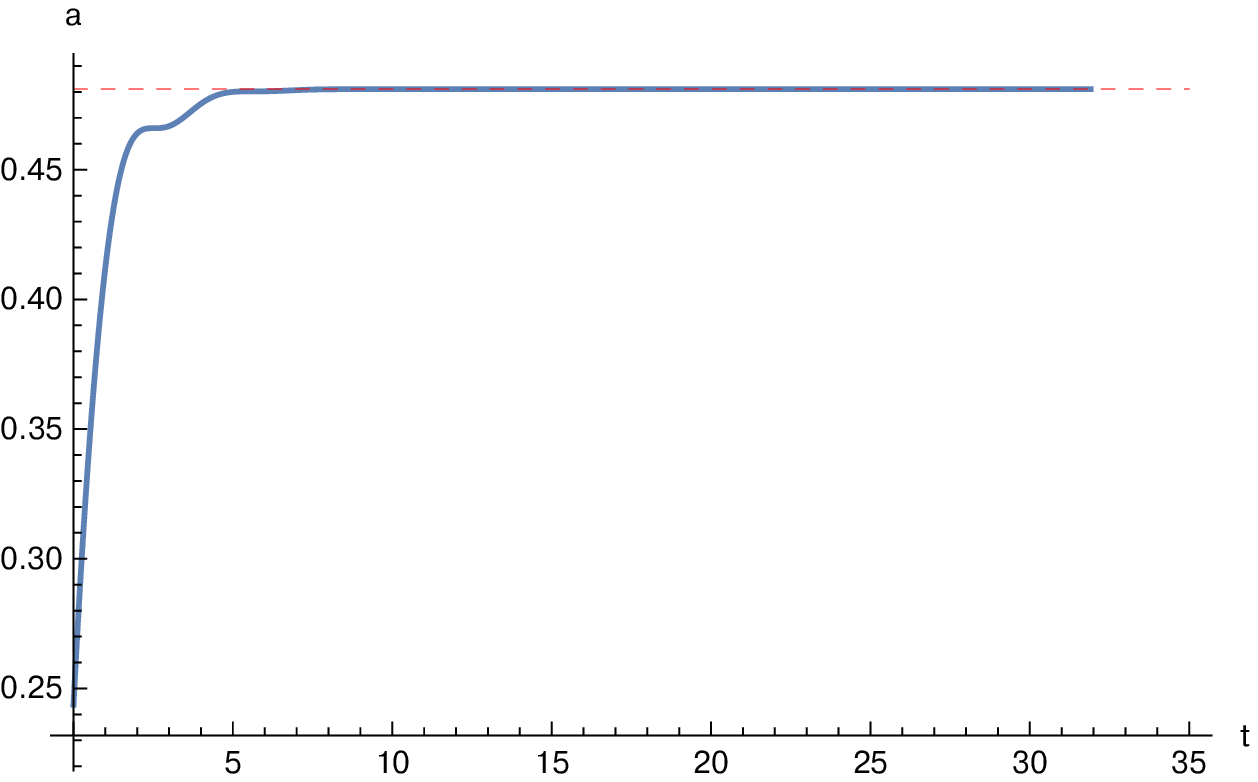}\,\,\,\,\,\,\,\,\,
  \includegraphics[width=2.5in]{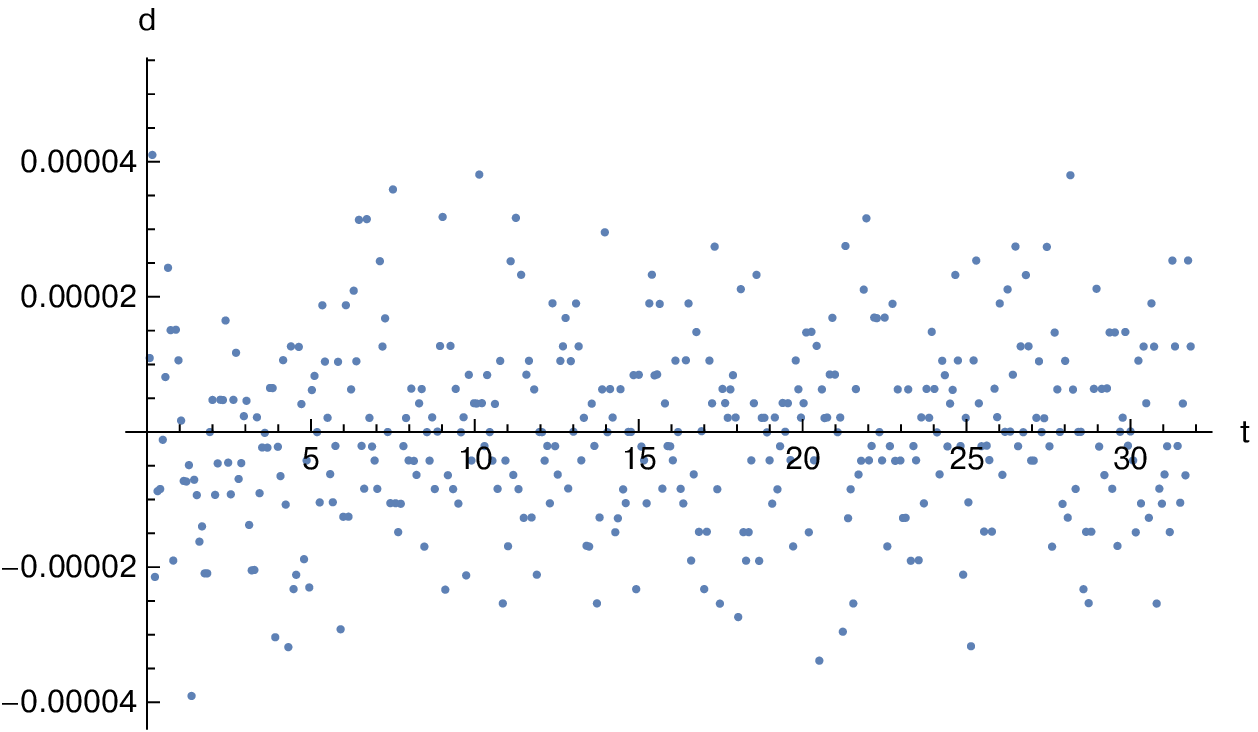}
\end{center}
 \caption{Evolution of the entropy density in  $\zet_2$ symmetric sector of sBBL model. 
The red dashed line indicates the equilibrium value (left panel). The entropy production rate 
is always positive during the evolution, see \eqref{srate}. The right panel 
presents numerical error in  computation of the entropy production rate, see \eqref{serr}. } \label{figure11}
\end{figure}

\begin{figure}[t]
\begin{center}
\psfrag{t}{{$t\Lambda$}}
\psfrag{q}{{$\ln|\langle\calo_i\rangle/\Lambda^4|$}}
  \includegraphics[width=2.5in]{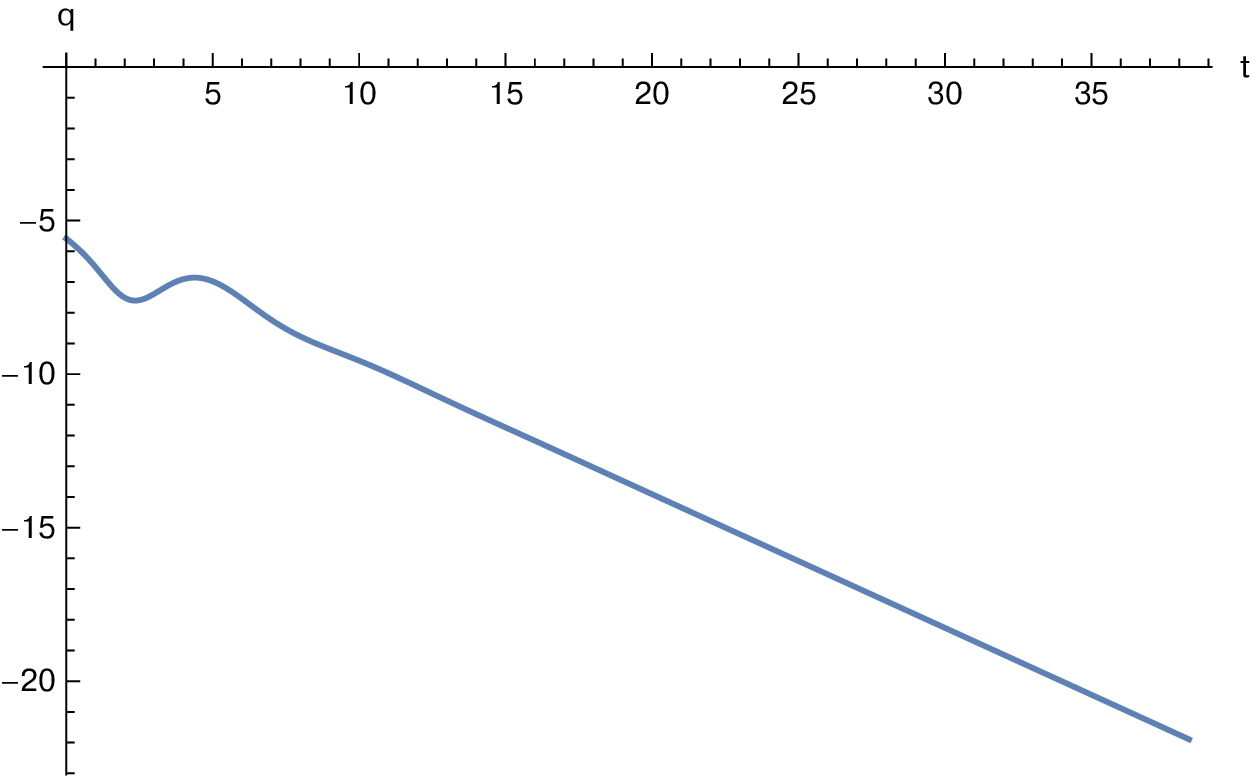}\,\,\,\,\,\,\,\,\,
  \includegraphics[width=2.5in]{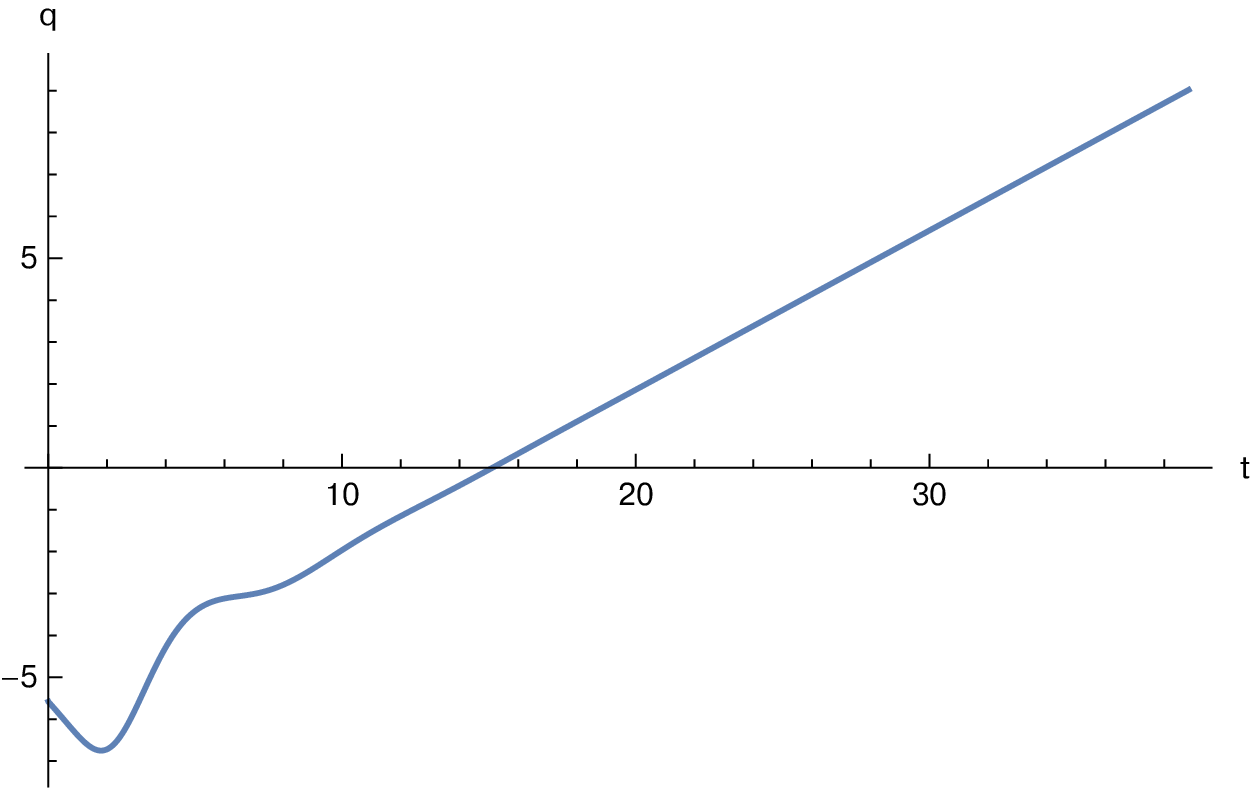}
\end{center}
 \caption{Linearized dynamics of the symmetry breaking expectation value 
$\langle\calo_i\rangle$ for $\cale<\cale_{crit}$ (left panel) and $\cale>\cale_{crit}$ 
(right panel).} \label{figure12}
\end{figure} 

Typical evolution in the symmetric sector (the bulk scalar $\chi$ identically vanishes) is shown in fig.~\ref{figure10}.
An important check on the consistency of the evolution is the fact that the expectation value of $\calo_r$ 
evolves at asymptotically late times to its equilibrium value at the corresponding energy density
(here we take $g=-8$ and $\cale=0.42978(1) \cale_{crit}$),  computed independently in section \ref{msbbl}.  

One of the advantages of the holographic formulation of the dynamics of strongly interactive gauge theories is 
a natural  definition of the non-equilibrium entropy density $s(t)$, associated with the area density of the apparent horizon,
\begin{equation}
s(t)=\frac{2\pi}{\k^2}\ \Sigma(t,r)^2\bigg|_{r=1}\,.
\eqlabel{sdyn}
\end{equation}
The evolution of the entropy density with time for the same set of parameters as in fig.~\ref{figure10}
is shown in fig.~\ref{figure11}. The red dashed line (left panel) 
indicates the equilibrium value of the entropy density
at the corresponding energy density, computed in  section \ref{msbbl}. 
From the plot it is clear that the entropy production rate is non-negative, \ie $\dot{s}_{sym}\ge 0$.
In fact, using the gravitational 
bulk equations of motion we find  
\begin{equation}
\dot{s}(t)= \frac{\k^2}{2\pi}\ \left(\Sigma^2\right)'\ \frac{(d_+\phi)^2+(d_+\chi)^2}{-2 \calp_{sBBL}}\bigg|_{r=1}\,,
\eqlabel{srate}
\end{equation} 
which can be analytically proven to be non-negative following \cite{Buchel:2017pto}.
Because a derivation of \eqref{srate} involves gravitational bulk constraint equations 
not directly used in the numerical evolution code, an important (dynamical) consistency check on the code 
is the vanishing of
\begin{equation}
\dd \dot{s} \equiv  \frac{\k^2}{2\pi}\  \left[ \frac {d}{dt} \Sigma^2
- \left(\Sigma^2\right)'\ \frac{(d_+\phi)^2+(d_+\chi)^2}{-2 \calp_{sBBL}}\right]\bigg|_{r=1}\,.
\eqlabel{serr}    
\end{equation}
The right panel in fig.~\ref{figure11} monitors the quantity \eqref{serr}. 

Taking the value of the coupling $g=-8$ in the bulk scalar potential, we study next the linearized dynamics of the 
$\zet_2$ symmetry breaking fluctuations.  Fig.\ref{figure12} presents the evolution of $\langle\calo_i\rangle$
without the backreaction on the symmetric sector dynamics. Once the symmetric sector 
equilibrates, $t\Lambda \gtrsim 10$, the linearize $\zet_2$ breaking fluctuations decay 
(left panel: $\cale=1.7191(2)\cale_{crit}$)
or grow (right panel: $\cale=0.42978(1)\cale_{crit}$) exponentially with time. 
Using the linear fit, we compare the decay/growth rates with the QNM spectrum predictions, 
see fig.~\ref{figure8}:
\begin{equation}
\bigg|\frac{\Im(\w)_{fit}}{\Im(\w)_{QNM}}-1\bigg| \lesssim \begin{cases} 10^{-5}\,,\qquad \cale>\cale_{crit}\\
10^{-6}\,,\qquad \cale<\cale_{crit}
\end{cases}\,.
\eqlabel{comp}
\end{equation}

\subsubsection{Unstable sBBL dynamics}

Having reproduced the phase diagram and the linearized symmetry breaking 
dynamics of sBBL model in section \ref{ldynsbbl}, we now present fully nonlinear
dynamical results. We focus on the unstable case only, as for simulations with $\cale>\cale_{crit}$,
after a brief non-linear regime, the system evolves to symmetric equilibrium configurations 
discussed above.

\begin{figure}[t]
\begin{center}
\psfrag{t}{{$t\Lambda$}}
\psfrag{i}{{$\ln|\langle\calo_i\rangle/\Lambda^4|$}}
\psfrag{r}{{$\langle\calo_r\rangle/\Lambda$}}
  \includegraphics[width=2.5in]{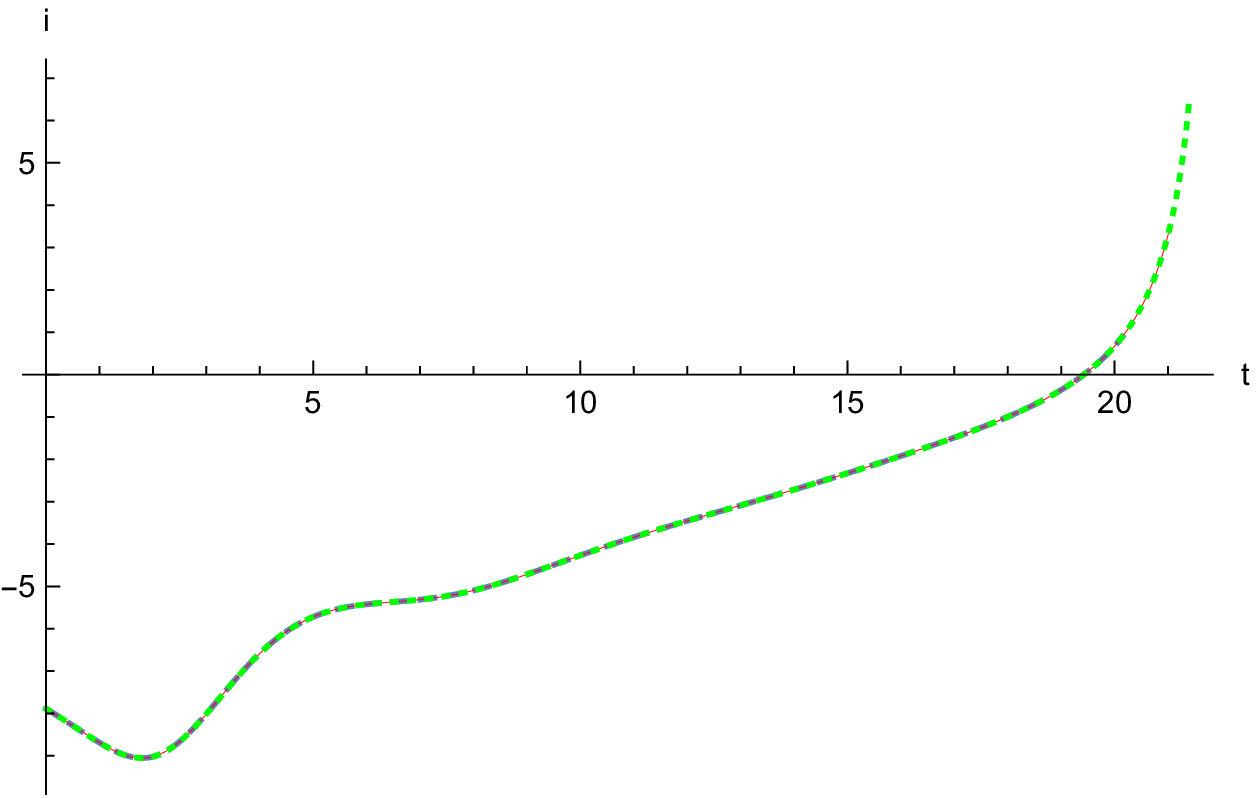}\,\,\,\,\,\,\,\,\,
  \includegraphics[width=2.5in]{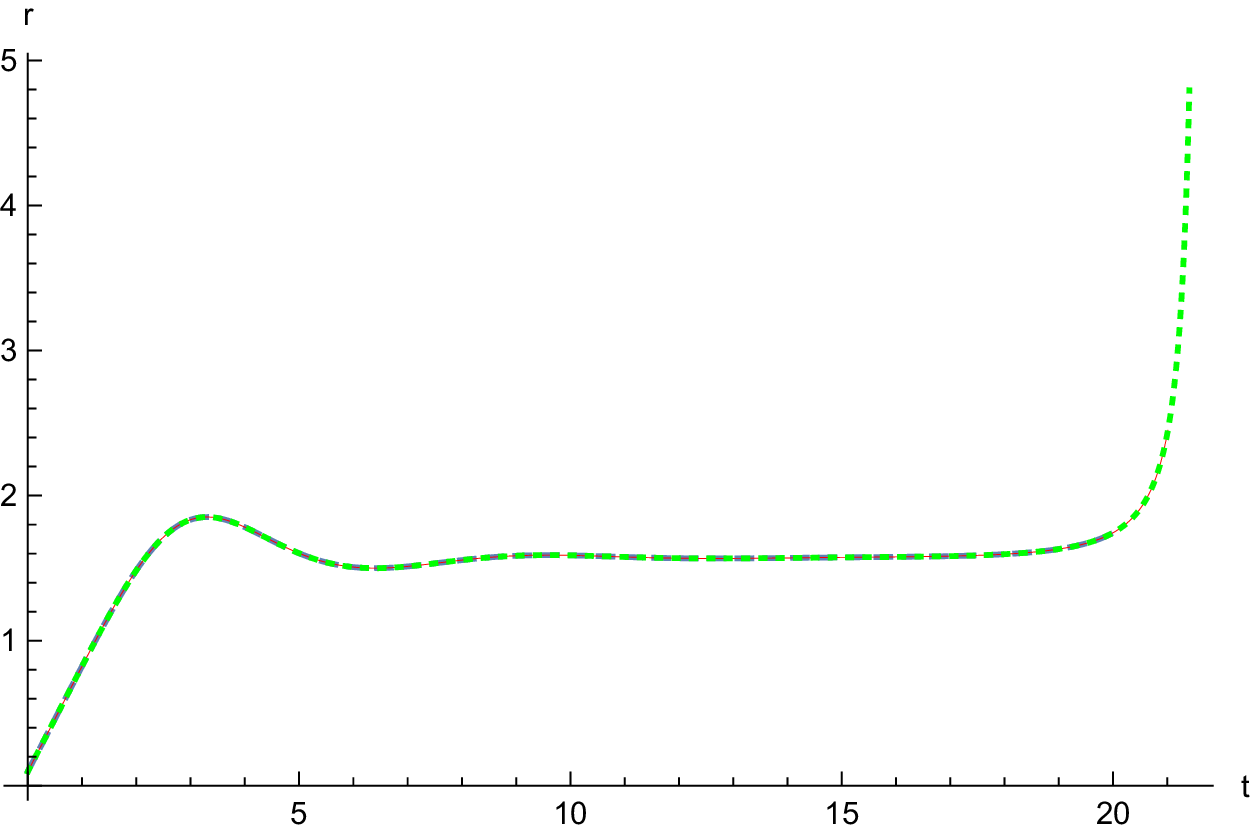}
\end{center}
 \caption{Evolution of expectation values of $\calo_i$ and $\calo_r$ operators in sBBL model 
with $g=-8$ in the unstable regime, \ie with $\cale<\cale_{crit}$. Different color coding represents 
different spatial resolutions of the numerical runs. }\label{figure13}
\end{figure}

\begin{figure}[t]
\begin{center}
\psfrag{t}{{$t\Lambda$}}
\psfrag{s}{{$\k^2 s/(2\pi \Lambda^2)$}}
\psfrag{k}{{$K^h/K_{AdS_4}$}}
  \includegraphics[width=2.5in]{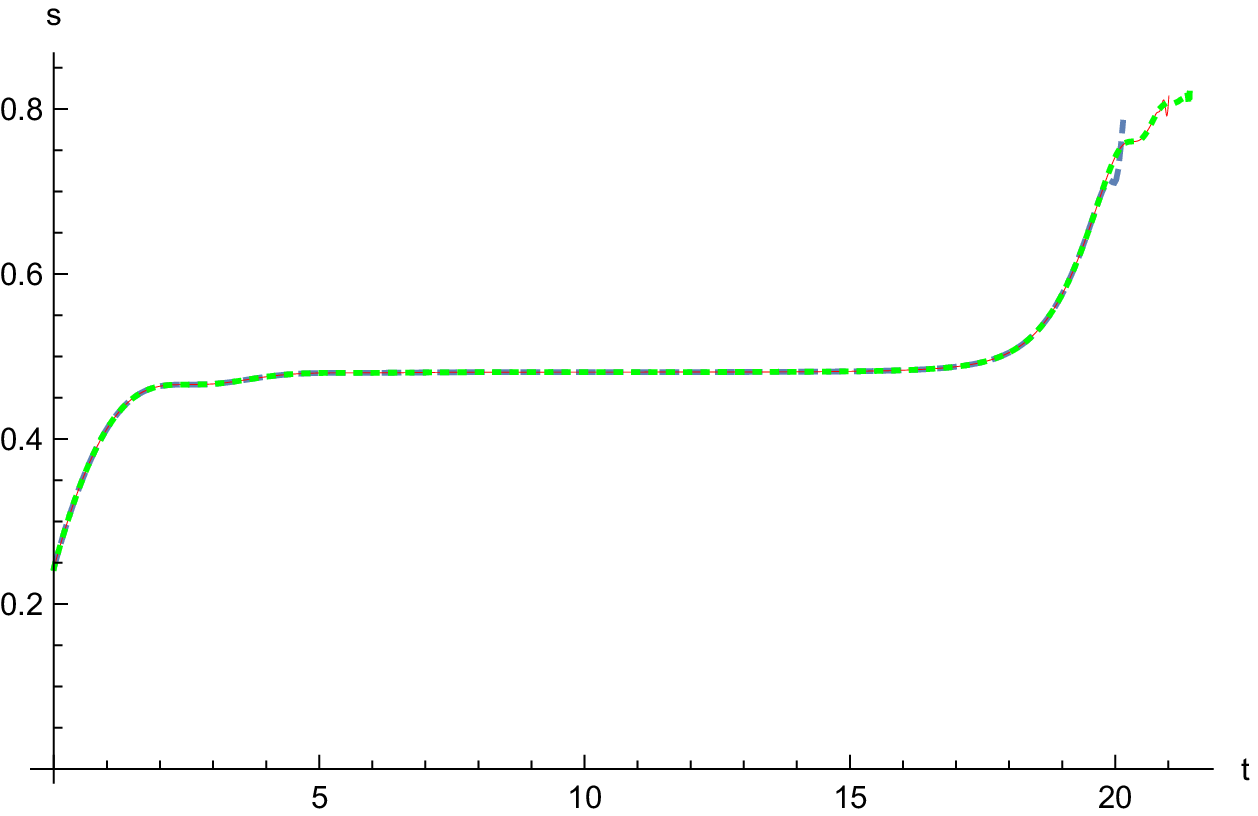}\,\,\,\,\,\,\,\,\,
  \includegraphics[width=2.5in]{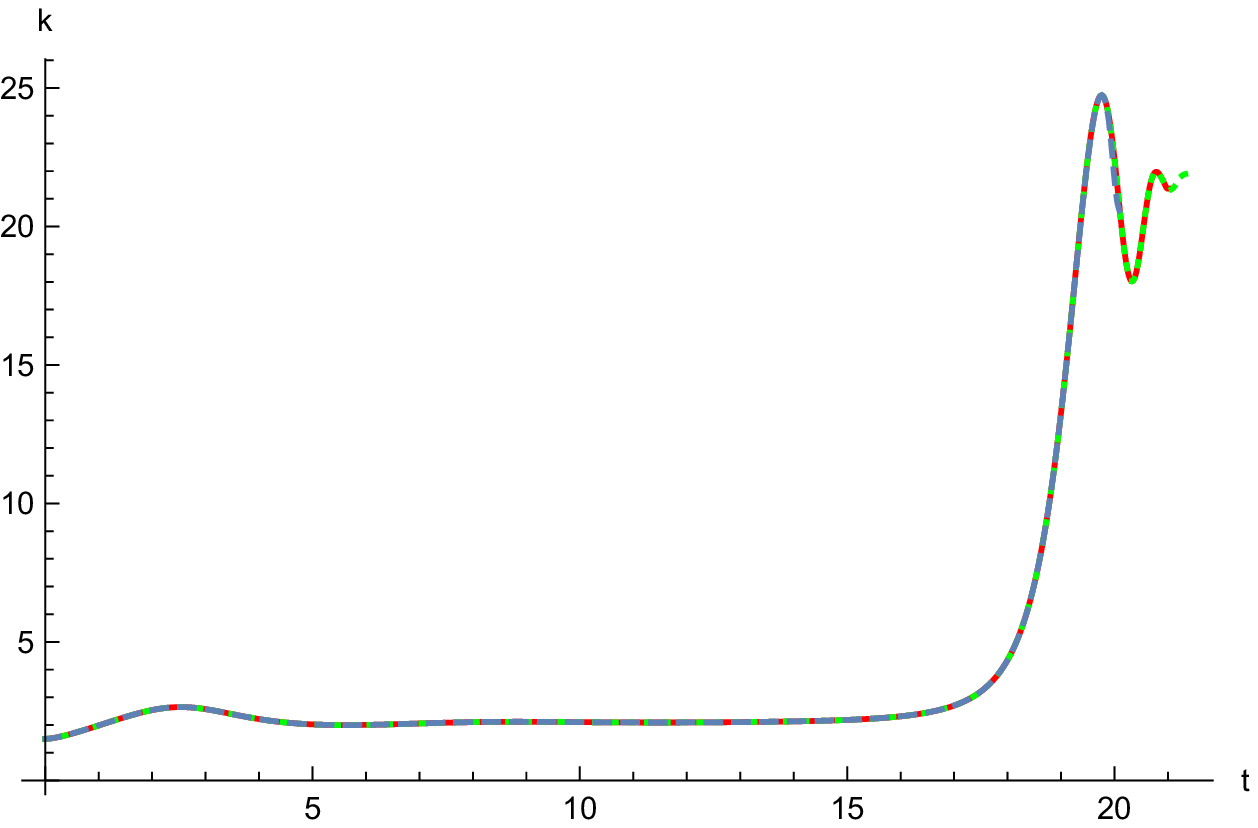}
\end{center}
 \caption{Evolution of the entropy density $s$ and the Kretschmann scalar $K^h$  evaluated at the apparent 
horizon in dynamics of sBBL model with  $g=-8$ in the unstable regime, \ie with $\cale<\cale_{crit}$. Different color coding represents 
different spatial resolutions of the numerical runs.} \label{figure14}
\end{figure}

We consider first the evolution with $g=-8$ and $\cale=0.42978(1)\cale_{crit}$ and $\cale_{crit}$ as in \eqref{ecrit28}. Recall that 
for this value of the coupling $\cale_{crit}>0$. Figs.~\ref{figure13}-\ref{figure14} collect results for the 
evolution of the expectation values of operators $\calo_i$, $\calo_r$, the entropy density $s$ and the Kretschmann scalar $K^h$  evaluated at the apparent 
horizon,
\begin{equation}
K^h=R_{abcd}R^{abcd}\bigg|_{(t,r=1)}\;,
\eqlabel{defK}
\end{equation}
relative to the $AdS_4$  Kretschmann scalar $K_{AdS_4}$ (recall $K_{AdS_4}={\rm const}=24$).
Different color coding on the plots represents numerical runs from the same initial conditions, but with different 
spatial relation  --- the different number of collocation points:
\begin{equation}
N_{collocation}=\begin{cases}
20\,,\qquad {\rm blue\ dashed}\\
40\,,\qquad {\rm red}\\
60\,,\qquad {\rm green\ dotted}
\end{cases}\,.
\eqlabel{ncol}
\end{equation}
As for the BBL model in \cite{Bosch:2017ccw}, the numerical code always crashes, albeit at a slightly later time with increasing spatial resolution. 
Similar to BBL model, instability in sBBL model persists in a nonlinear regime --- we do not see the saturation in expectation values of 
$\calo_i$ and $\calo_r$. However, contrary to BBL, in sBBL model with $g=-8$ ($\cale_{crit}>0$) we do not see a clear signature of the divergence 
of the area density of the apparent horizon (the entropy density); moreover, there is no obvious 
divergence in  the Kretschmann scalar $K^h$ as well. 
Clearly, a better (different) implementation\footnote{A finite-difference code as in \cite{Buchel:2012gw}
might be more suitable to capture high-gradients in the evolution. Even 
more leverage could be achieved with adaptive mesh refinement as in \cite{Buchel:2012uh}.} of the code is needed to answer conclusively whether
 the entropy density and/or  the Kretschmann scalar
remain finite.

\begin{figure}[t]
\begin{center}
\psfrag{t}{{$t\Lambda$}}
\psfrag{i}{{$\ln|\langle\calo_i\rangle/\Lambda^4|$}}
\psfrag{r}{{$\langle\calo_r\rangle/\Lambda$}}
  \includegraphics[width=2.5in]{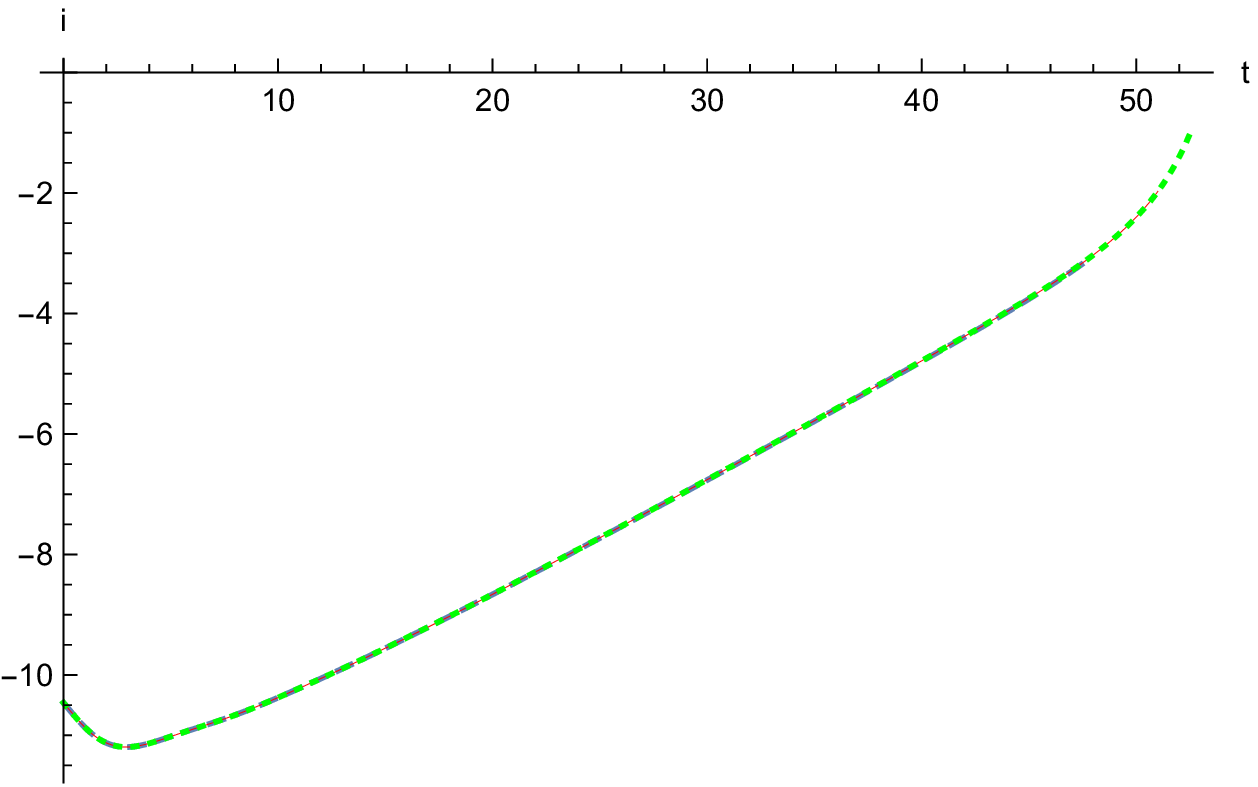}\,\,\,\,\,\,\,\,\,
  \includegraphics[width=2.5in]{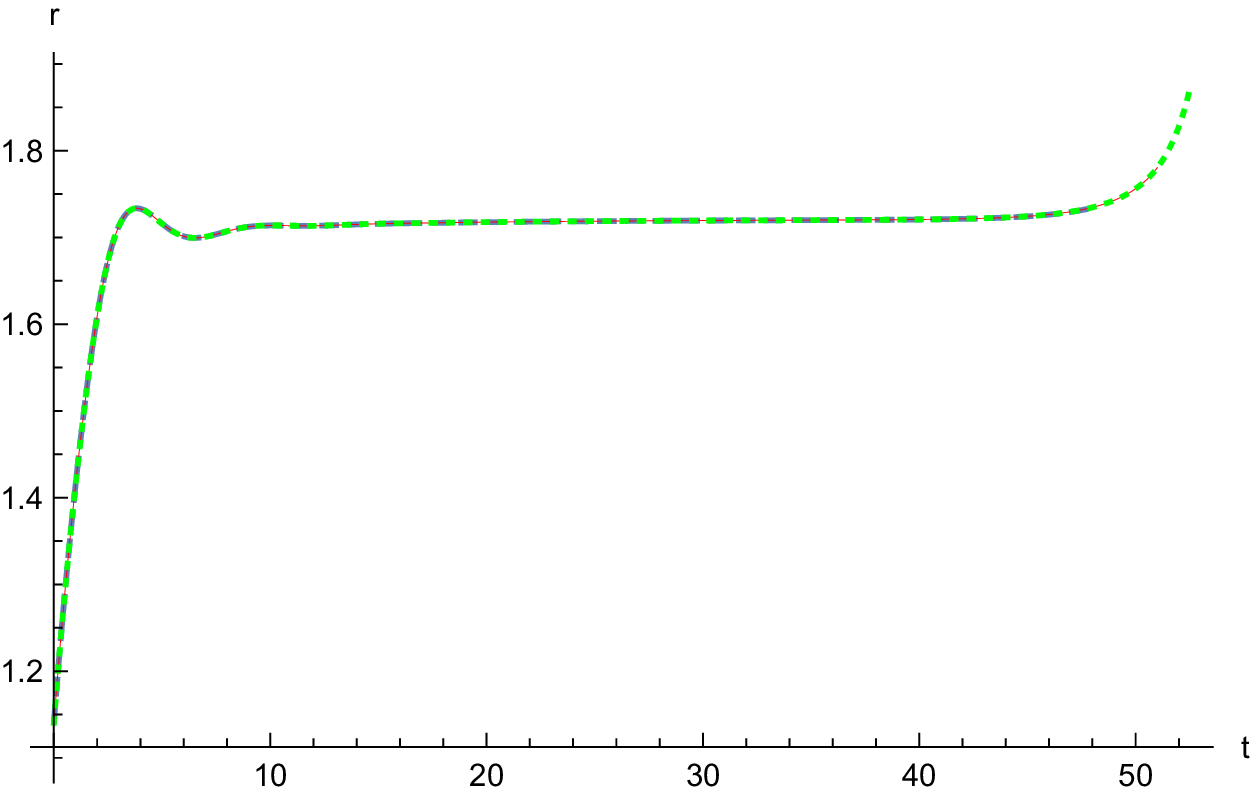}
\end{center}
 \caption{Evolution of expectation values of $\calo_i$ and $\calo_r$ operators in sBBL model 
with $g=-2$ in the unstable regime, \ie with $\cale<\cale_{crit}$. Different color coding represents 
different spatial resolutions of the numerical runs. } \label{figure15}
\end{figure}

\begin{figure}[t]
\begin{center}
\psfrag{t}{{$t\Lambda$}}
\psfrag{s}{{$\k^2 s/(2\pi \Lambda^2)$}}
\psfrag{k}{{$K^h/K_{AdS_4}$}}
  \includegraphics[width=2.5in]{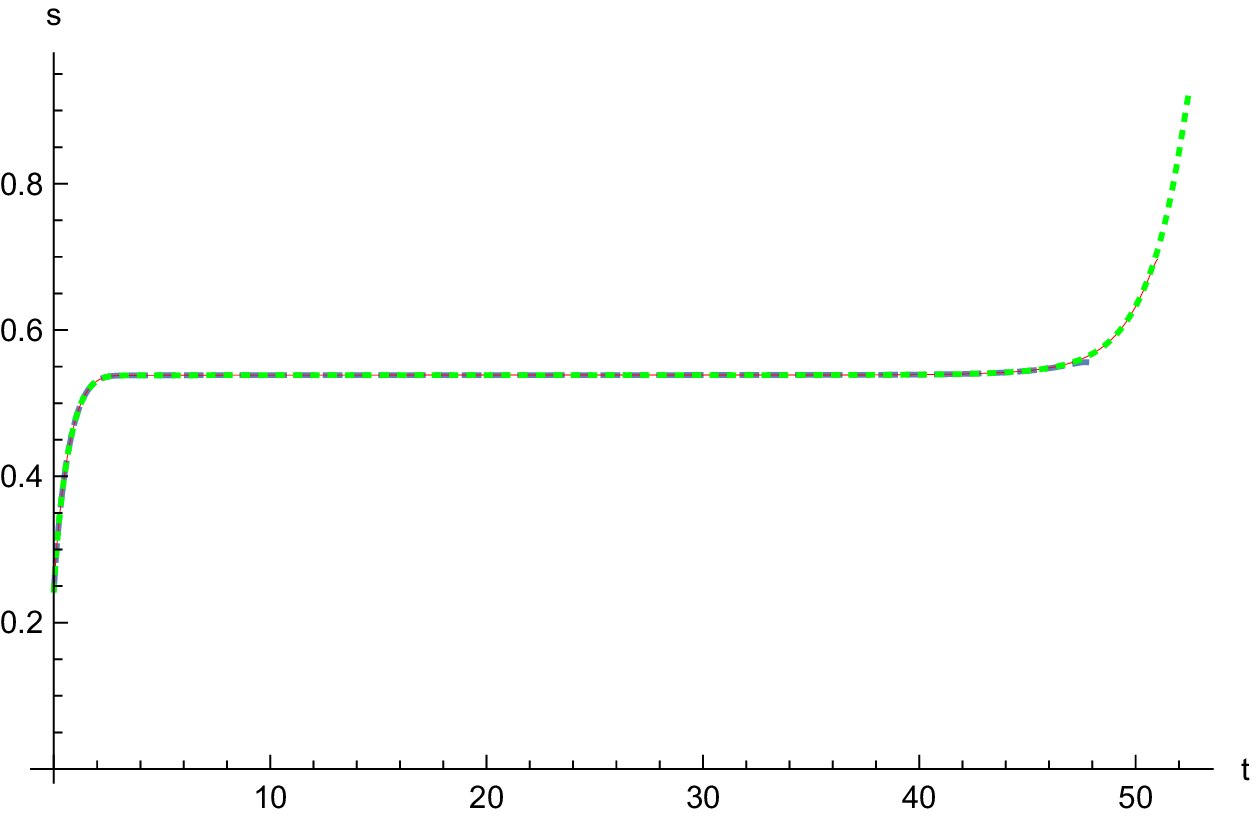}\,\,\,\,\,\,\,\,\,
  \includegraphics[width=2.5in]{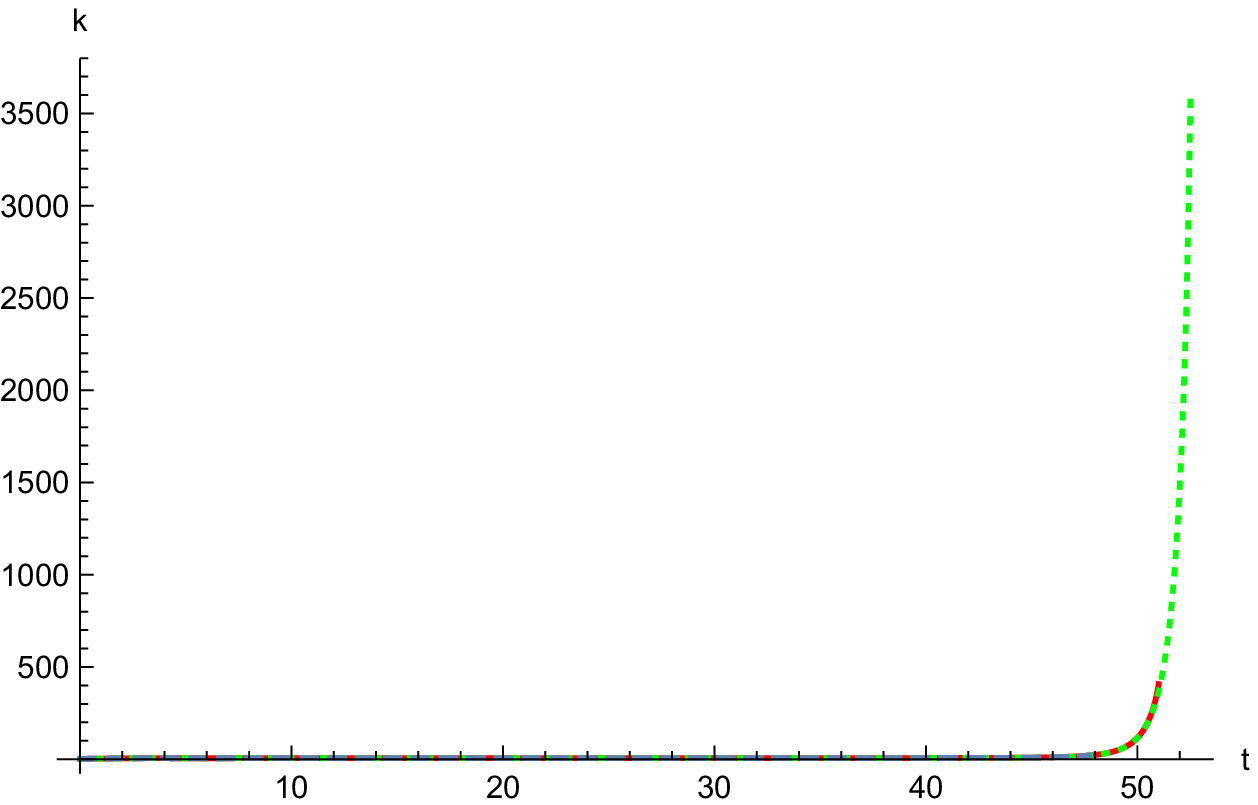}
\end{center}
 \caption{Evolution of the entropy density $s$ and the Kretschmann scalar $K^h$  evaluated at the apparent 
horizon  in dynamics of sBBL model with  $g=-2$ in the unstable regime, \ie with $\cale<\cale_{crit}$. Different color coding represents 
different spatial resolutions of the numerical runs.} \label{figure16}
\end{figure}

Figs.~\ref{figure15}-\ref{figure16} collect the same data, but for the sBBL coupling constant $g=-2$ and 
$\cale=1.0454(8)\cale_{crit}$ with the negative critical energy as in \eqref{ecrit28}. Here we have 
a stronger indication for the divergence of the Kretschmann scalar $K^h$ along with the more dramatic growth 
of the entropy density $s$. Better implementation of the numerics could answer whether dynamics in sBBL model 
at $g=-2$ is physically different from that at $g=-8$, and whether this difference is attributed to the sign of $\cale_{crit}$.

\section{Conclusions}\label{conclude}

In \cite{Bosch:2017ccw} it was argued that the phenomenological holographic model  introduced in \cite{Buchel:2009ge} 
violates the weak cosmic censorship conjecture. The gravitational dual describes dynamics of certain $QFT_3$ with spontaneous symmetry breaking,
but without an equilibrium ground state below the instability threshold\footnote{Potentially related phenomenon was 
reported in \cite{Gursoy:2016ggq}.}. An arbitrary weakly curved initial gravitational configuration 
was shown to evolve, in a finite boundary time, to a configuration with both the divergent area density of the apparent horizon and the 
Kretschmann scalar evaluated at the horizon. The divergence of the area density signals that the singularity mechanism is robust --- while the evolution 
was restricted to  spatially homogeneous and isotropic states of the $QFT_3$, any finite entropy density  state with broken (boundary) 
spatial translational invariance and/or rotations can not dominate late time dynamics.  Following the gravity-fluid correspondence 
\cite{Bhattacharyya:2008jc}, the BBL holographic model points to a development of the singularities from regular initial conditions 
in corresponding relativistic fluid mechanics. 

In this paper we attempted to address two questions:
\nxt (1): can BBL scenario be realized in a top-down string construction?
\nxt (2): what is the role of supersymmetry on the singularity development? 

Concerning (1), it was pointed out in \cite{Bosch:2017ccw} that consistent truncations studied in \cite{Donos:2011ut}
exhibit the exotic thermodynamics of  \cite{Buchel:2009ge}  in grand canonical ensemble. We studied here the corresponding 
models in details and established that there is no run-away instability in DG models in microcanonical ensemble --- rather, 
we found yet another realization of the mean-field spontaneous symmetry breaking mechanism of \cite{Hartnoll:2008vx,Hartnoll:2008kx}.

Concerning (2), we showed that it is straightforward to modify BBL model to mimic the structure of the bulk scalar coupling to gravity 
ubiquitous in gauged supergravity consistent truncations  \cite{Ahn:2000mf,Bobev:2009ms}. The scalar potential of the "supersymmetric'' generalization of the 
BBL model (called sBBL here) remains unbounded from below. The equilibrium phase diagram and the linearized symmetry breaking dynamics in BBL and sBBL 
models are conceptually identical. Further studies (and a better numerical code) are needed to firmly establish whether sBBL model also evolves 
to a geometry with divergent area density of the apparent horizon and the curvature. As we pointed out, the latter might depend on details 
of the scalar superpotential, \ie the nonlinear coupling $g$. 

Finally, the challenge remains to find embedding of BBL mechanism in string theory.

\appendix
\section{Numerical setup}\label{appendix}

We adapt numerical code developed in \cite{Bosch:2017ccw} to study 
dynamics of spatially homogeneous and isotropic states in DG and sBBL models.

\subsection{DG-B model}\label{dgbnum}

We introduce a new radial coordinate  
\begin{equation}
x\equiv \frac 1r\in [0,1]\,,\qquad d_+ =\del_t +A(t,r)\ \del_r \ \to\ \del_t-x^2 A(t,x) \ \del_x\,,  
\eqlabel{redef}
\end{equation}
maintaining $'\equiv \del_x$ and $\dot\ \equiv \del_t$,
and redefine the fields 
\begin{equation}
\{\, \Sigma\,,\, A\,,\, d_+\Sigma\, \}
\ \to\
\{\, \sigma\,,\, a\,,\, d\sigma\, \}\,,
\eqlabel{fieldsor}
\end{equation}
as follows
\begin{equation}
\begin{split}
&\Sigma(t,x)=\frac 1x+\sigma(t,x)\;,\\
&A=a(t,x)-\Sigma(t,x)^2+\frac 2x\ \Sigma(t,x)+2\l(t)\ \Sigma(t,x)\;,\\
&d_+\Sigma(t,x)=x\ \ds(t,x)+\Sigma(t,x)^2-\frac 1x\ \Sigma(t,x)+\frac{1+x\ \l(t)}{x^2}\,.
\end{split}
\eqlabel{fieldnew}
\end{equation}
Using \eqref{uvdgb} and \eqref{fi2}, 
we find the asymptotic boundary expansion $x\to 0_+$ for all the  fields:
\begin{equation}
\begin{split}
&f_i=f_{i,1}(t)\ x + \calo(x^2)\,,\qquad d_+f_i=-f_{i,1}(t)+\calo(x^2)\,,\\
&\sigma=\l(t)-\frac{1}{2}\left(f_{1,1}(t)^2+f_{2,1}(t)^2\right)\ x+\calo(x^2)\,,\\  
&\ds=m+\frac{\lambda(t)}{2}\left(f_{1,1}(t)^2+f_{2,1}(t)^2\right)+\calo(x)\,,\\
&a=-\dot\l(t)-f_{1,1}(t)^2-f_{2,1}(t)^2+ m\ x+\calo(x^2)\,,\\
&a_1= q_1\ x + \calo(x^2)\,,
\end{split}
\eqlabel{newbasym}
\end{equation}
where we set 
\begin{equation}
\{A_1(t), a_{1,1}(t)\} = \{m, q_1\}\,.
\eqlabel{seta1a11}
\end{equation}
Note that the boundary conditions \eqref{fi2} are enforced with the absence of $\calo(x)$ terms in 
asymptotic expansion of $d_+f_i$ in \eqref{newbasym}. 

The rest of the code implementation is as in \cite{Bosch:2017ccw}.

\subsection{sBBL model}\label{numsbbl} 

We use a radial coordinate $x$ as in \eqref{redef} and employ the following field redefinitions:
\begin{equation}
\{\, \phi\,,\, \chi\,,\, \Sigma\,,\, A\,,\, d_+\phi\,,\, d_+\chi\,,\, d_+\Sigma\, \}
\ \to\
\{\, p\,,\, q\,,\, \sigma\,,\, a\,,\, dp\,,\, dq\,,\, d\sigma\, \}\,,
\eqlabel{fieldsor2}
\end{equation}
with
\begin{equation}
\begin{split}
&\phi(t,x)=p(t,x)\;,\\
&\chi(t,x)=x^3\ q(t,x)\;,\\
&\Sigma(t,x)=\frac 1x+\sigma(t,x)\;,\\
&A=a(t,x)+\frac 12\ \Sigma(t,x)^2-\frac 1x\ \Sigma(t,x)+\frac{1+x\l(t)}{x^2}\;,\\
&d_+\phi(t,x)=dp(t,x)\;,\\
&d_+\chi(t,x)=x^3\ dq(t,x)\;,\\
&d_+\Sigma(t,x)=x\ \ds(t,x)+\frac 12\ \Sigma(t,x)^2-\frac{1}{2x}\ \Sigma(t,x)+\frac{1+x\l(t)}{2x^2}\,.
\end{split}
\eqlabel{fieldnew2}
\end{equation}
Using \eqref{uvsbbl}, we find the asymptotic boundary expansion $x\to 0_+$ for the new fields:
\begin{equation}
\begin{split}
&p=p_1(t)\ x + \calo(x^2)\,,\qquad q=q_4(t)\ x+\calo(x^2)\;,\\
&dp=-\frac{p_1(t)}{2} -p_2\ x+\calo(x^2)\,,\qquad dq=-2 q_4(t)+\calo(x)\;,\\
&\sigma=\l(t)-\frac{p_1(t)^2}{8}\ x+\calo(x^2)\,,\qquad \ds=\frac{\l(t) p_1(t)^2}{16}+\frac{p_1(t)p_2}{4}
+\mu+\calo(x)\;,\\
&a=-\frac{p_1(t)^2}{8}-\dot\l(t)+\left(\mu+\frac 18 \l(t) p_1(t)^2\right)\
x+\calo(x^2)\;.
\end{split}
\eqlabel{newbasym2}
\end{equation}

The rest of the code implementation is as in \cite{Bosch:2017ccw}.

\section*{Acknowledgments}
Research at Perimeter
Institute is supported by the Government of Canada through Industry
Canada and by the Province of Ontario through the Ministry of
Research \& Innovation. This work was further supported by
NSERC through the Discovery Grants program.

%\appendix
%\section{Apparent horizon area growth theorem}\label{th1}


\begin{thebibliography}{99}


%\cite{Bosch:2017ccw}
\bibitem{Bosch:2017ccw} 
  P.~Bosch, A.~Buchel and L.~Lehner,
  ``Unstable horizons and singularity development in holography,''
  arXiv:1704.05454 [hep-th].
  %%CITATION = ARXIV:1704.05454;%%


\bibitem %[m1]
 {m1} J.~M.~Maldacena,
  ``The large N limit of superconformal field theories and supergravity,''
  Adv.\ Theor.\ Math.\ Phys.\  {\bf 2}, 231 (1998)
  [Int.\ J.\ Theor.\ Phys.\  {\bf 38}, 1113 (1999)]
  [arXiv:hep-th/9711200].


%\cite{Aharony:1999ti}
\bibitem{Aharony:1999ti} 
  O.~Aharony, S.~S.~Gubser, J.~M.~Maldacena, H.~Ooguri and Y.~Oz,
  ``Large N field theories, string theory and gravity,''
  Phys.\ Rept.\  {\bf 323}, 183 (2000)
  [hep-th/9905111].
  %%CITATION = HEP-TH/9905111;%%
  %3348 citations counted in INSPIRE as of 05 Jan 2015

%\cite{Witten:1998zw}
\bibitem{Witten:1998zw} 
  E.~Witten,
  ``Anti-de Sitter space, thermal phase transition, and confinement in gauge theories,''
  Adv.\ Theor.\ Math.\ Phys.\  {\bf 2}, 505 (1998)
  [hep-th/9803131].
  %%CITATION = HEP-TH/9803131;%%
  %2547 citations counted in INSPIRE as of 15 May 2017

%\cite{Bhattacharyya:2008jc}
\bibitem{Bhattacharyya:2008jc} 
  S.~Bhattacharyya, V.~E.~Hubeny, S.~Minwalla and M.~Rangamani,
  ``Nonlinear Fluid Dynamics from Gravity,''
  JHEP {\bf 0802}, 045 (2008)
  doi:10.1088/1126-6708/2008/02/045
  [arXiv:0712.2456 [hep-th]].
  %%CITATION = doi:10.1088/1126-6708/2008/02/045;%%
  %687 citations counted in INSPIRE as of 15 May 2017

%\cite{Heller:2013fn}
\bibitem{Heller:2013fn} 
  M.~P.~Heller, R.~A.~Janik and P.~Witaszczyk,
  ``Hydrodynamic Gradient Expansion in Gauge Theory Plasmas,''
  Phys.\ Rev.\ Lett.\  {\bf 110}, no. 21, 211602 (2013)
  doi:10.1103/PhysRevLett.110.211602
  [arXiv:1302.0697 [hep-th]].
  %%CITATION = doi:10.1103/PhysRevLett.110.211602;%%
  %63 citations counted in INSPIRE as of 15 May 2017


%\cite{Buchel:2016cbj}
\bibitem{Buchel:2016cbj} 
  A.~Buchel, M.~P.~Heller and J.~Noronha,
  ``Entropy Production, Hydrodynamics, and Resurgence in the Primordial Quark-Gluon Plasma from Holography,''
  Phys.\ Rev.\ D {\bf 94}, no. 10, 106011 (2016)
  doi:10.1103/PhysRevD.94.106011
  [arXiv:1603.05344 [hep-th]].
  %%CITATION = doi:10.1103/PhysRevD.94.106011;%%
  %13 citations counted in INSPIRE as of 15 May 2017


%\cite{Berti:2009kk}
\bibitem{Berti:2009kk} 
  E.~Berti, V.~Cardoso and A.~O.~Starinets,
  ``Quasinormal modes of black holes and black branes,''
  Class.\ Quant.\ Grav.\  {\bf 26}, 163001 (2009)
  doi:10.1088/0264-9381/26/16/163001
  [arXiv:0905.2975 [gr-qc]].
  %%CITATION = doi:10.1088/0264-9381/26/16/163001;%%
  %565 citations counted in INSPIRE as of 15 May 2017


%\cite{Buchel:2015saa}
\bibitem{Buchel:2015saa} 
  A.~Buchel, M.~P.~Heller and R.~C.~Myers,
  ``Equilibration rates in a strongly coupled nonconformal quark-gluon plasma,''
  Phys.\ Rev.\ Lett.\  {\bf 114}, no. 25, 251601 (2015)
  doi:10.1103/PhysRevLett.114.251601
  [arXiv:1503.07114 [hep-th]].
  %%CITATION = doi:10.1103/PhysRevLett.114.251601;%%
  %29 citations counted in INSPIRE as of 15 May 2017


%\cite{Fuini:2015hba}
\bibitem{Fuini:2015hba} 
  J.~F.~Fuini and L.~G.~Yaffe,
  ``Far-from-equilibrium dynamics of a strongly coupled non-Abelian plasma with non-zero charge density or external magnetic field,''
  JHEP {\bf 1507}, 116 (2015)
  doi:10.1007/JHEP07(2015)116
  [arXiv:1503.07148 [hep-th]].
  %%CITATION = doi:10.1007/JHEP07(2015)116;%%
  %23 citations counted in INSPIRE as of 15 May 2017


%\cite{Janik:2015waa}
\bibitem{Janik:2015waa} 
  R.~A.~Janik, G.~Plewa, H.~Soltanpanahi and M.~Spalinski,
  ``Linearized nonequilibrium dynamics in nonconformal plasma,''
  Phys.\ Rev.\ D {\bf 91}, no. 12, 126013 (2015)
  doi:10.1103/PhysRevD.91.126013
  [arXiv:1503.07149 [hep-th]].
  %%CITATION = doi:10.1103/PhysRevD.91.126013;%%
  %25 citations counted in INSPIRE as of 15 May 2017


%\cite{Buchel:2015ofa}
\bibitem{Buchel:2015ofa} 
  A.~Buchel and A.~Day,
  ``Universal relaxation in quark-gluon plasma at strong coupling,''
  Phys.\ Rev.\ D {\bf 92}, no. 2, 026009 (2015)
  doi:10.1103/PhysRevD.92.026009
  [arXiv:1505.05012 [hep-th]].
  %%CITATION = doi:10.1103/PhysRevD.92.026009;%%
  %7 citations counted in INSPIRE as of 15 May 2017

%\cite{Buchel:2005nt}
\bibitem{Buchel:2005nt} 
  A.~Buchel,
  ``A Holographic perspective on Gubser-Mitra conjecture,''
  Nucl.\ Phys.\ B {\bf 731}, 109 (2005)
  doi:10.1016/j.nuclphysb.2005.10.014
  [hep-th/0507275].
  %%CITATION = doi:10.1016/j.nuclphysb.2005.10.014;%%
  %31 citations counted in INSPIRE as of 15 May 2017

%\cite{Attems:2017ezz}
\bibitem{Attems:2017ezz} 
  M.~Attems, Y.~Bea, J.~Casalderrey-Solana, D.~Mateos, M.~Triana and M.~Zilhao,
  ``Phase Transitions, Inhomogeneous Horizons and Second-Order Hydrodynamics,''
  arXiv:1703.02948 [hep-th].
  %%CITATION = ARXIV:1703.02948;%%
  %3 citations counted in INSPIRE as of 15 May 2017


%\cite{Hartnoll:2008vx}
\bibitem{Hartnoll:2008vx} 
  S.~A.~Hartnoll, C.~P.~Herzog and G.~T.~Horowitz,
  ``Building a Holographic Superconductor,''
  Phys.\ Rev.\ Lett.\  {\bf 101}, 031601 (2008)
  doi:10.1103/PhysRevLett.101.031601
  [arXiv:0803.3295 [hep-th]].
  %%CITATION = doi:10.1103/PhysRevLett.101.031601;%%
  %989 citations counted in INSPIRE as of 15 May 2017



%\cite{Hartnoll:2008kx}
\bibitem{Hartnoll:2008kx} 
  S.~A.~Hartnoll, C.~P.~Herzog and G.~T.~Horowitz,
  ``Holographic Superconductors,''
  JHEP {\bf 0812}, 015 (2008)
  doi:10.1088/1126-6708/2008/12/015
  [arXiv:0810.1563 [hep-th]].
  %%CITATION = doi:10.1088/1126-6708/2008/12/015;%%
  %737 citations counted in INSPIRE as of 15 May 2017

%\cite{Buchel:2009ge}
\bibitem{Buchel:2009ge} 
  A.~Buchel and C.~Pagnutti,
  ``Exotic Hairy Black Holes,''
  Nucl.\ Phys.\ B {\bf 824}, 85 (2010)
  doi:10.1016/j.nuclphysb.2009.08.017
  [arXiv:0904.1716 [hep-th]].
  %%CITATION = doi:10.1016/j.nuclphysb.2009.08.017;%%
  %21 citations counted in INSPIRE as of 15 May 2017


%\cite{Buchel:2010wk}
\bibitem{Buchel:2010wk} 
  A.~Buchel and C.~Pagnutti,
  ``Correlated stability conjecture revisited,''
  Phys.\ Lett.\ B {\bf 697}, 168 (2011)
  doi:10.1016/j.physletb.2011.01.057
  [arXiv:1010.5748 [hep-th]].
  %%CITATION = doi:10.1016/j.physletb.2011.01.057;%%
  %10 citations counted in INSPIRE as of 15 May 2017

%\cite{Booth:2005qc}
\bibitem{Booth:2005qc} 
  I.~Booth,
  ``Black hole boundaries,''
  Can.\ J.\ Phys.\  {\bf 83}, 1073 (2005)
  doi:10.1139/p05-063
  [gr-qc/0508107].
  %%CITATION = doi:10.1139/p05-063;%%
%  %112 citations counted in INSPIRE as of 30 Jul 2016
%
%\cite{Figueras:2009iu}
\bibitem{Figueras:2009iu} 
  P.~Figueras, V.~E.~Hubeny, M.~Rangamani and S.~F.~Ross,
  ``Dynamical black holes and expanding plasmas,''
  JHEP {\bf 0904}, 137 (2009)
  doi:10.1088/1126-6708/2009/04/137
  [arXiv:0902.4696 [hep-th]].
%  %%CITATION = doi:10.1088/1126-6708/2009/04/137;%%
%  %55 citations counted in INSPIRE as of 30 Jul 2016

%%\cite{Donos:2011ut}
\bibitem{Donos:2011ut} 
  A.~Donos and J.~P.~Gauntlett,
  ``Superfluid black branes in $AdS_4\times S^7$,''
  JHEP {\bf 1106}, 053 (2011)
  doi:10.1007/JHEP06(2011)053
  [arXiv:1104.4478 [hep-th]].
%  %%CITATION = doi:10.1007/JHEP06(2011)053;%%
%  %38 citations counted in INSPIRE as of 22 Jun 2016
%


%\cite{Aharony:2007vg}
\bibitem{Aharony:2007vg} 
  O.~Aharony, A.~Buchel and P.~Kerner,
  %``The Black hole in the throat: Thermodynamics of strongly coupled cascading gauge theories,''
  Phys.\ Rev.\ D {\bf 76}, 086005 (2007)
  doi:10.1103/PhysRevD.76.086005
  [arXiv:0706.1768 [hep-th]].
  %%CITATION = doi:10.1103/PhysRevD.76.086005;%%
  %62 citations counted in INSPIRE as of 16 May 2017

\bibitem{Bosch:2016vcp} 
  P.~Bosch, S.~R.~Green and L.~Lehner,
  %``Nonlinear Evolution and Final Fate of Charged Anti–de Sitter Black Hole Superradiant Instability,''
  Phys.\ Rev.\ Lett.\  {\bf 116}, no. 14, 141102 (2016)
  doi:10.1103/PhysRevLett.116.141102
  [arXiv:1601.01384 [gr-qc]].
%  %%CITATION = doi:10.1103/PhysRevLett.116.141102;%%
%  %17 citations counted in INSPIRE as of 19 Dec 2016
%


%\cite{Ahn:2000mf}
\bibitem{Ahn:2000mf} 
  C.~h.~Ahn and K.~Woo,
  ``Supersymmetric domain wall and RG flow from 4-dimensional gauged N=8 supergravity,''
  Nucl.\ Phys.\ B {\bf 599}, 83 (2001)
  doi:10.1016/S0550-3213(01)00008-6
  [hep-th/0011121].
  %%CITATION = doi:10.1016/S0550-3213(01)00008-6;%%
  %45 citations counted in INSPIRE as of 22 May 2017

%\cite{Bobev:2009ms}
\bibitem{Bobev:2009ms} 
  N.~Bobev, N.~Halmagyi, K.~Pilch and N.~P.~Warner,
  ``Holographic, N=1 Supersymmetric RG Flows on M2 Branes,''
  JHEP {\bf 0909}, 043 (2009)
  doi:10.1088/1126-6708/2009/09/043
  [arXiv:0901.2736 [hep-th]].
  %%CITATION = doi:10.1088/1126-6708/2009/09/043;%%
  %40 citations counted in INSPIRE as of 22 May 2017

%\cite{Nunez:2003eq}
\bibitem{Nunez:2003eq} 
  A.~Nunez and A.~O.~Starinets,
  ``AdS / CFT correspondence, quasinormal modes, and thermal correlators in N=4 SYM,''
  Phys.\ Rev.\ D {\bf 67}, 124013 (2003)
  doi:10.1103/PhysRevD.67.124013
  [hep-th/0302026].
  %%CITATION = doi:10.1103/PhysRevD.67.124013;%%
  %142 citations counted in INSPIRE as of 23 May 2017

%\cite{Buchel:2017pto}
\bibitem{Buchel:2017pto} 
  A.~Buchel and A.~Karapetyan,
  ``de Sitter Vacua of Strongly Interacting QFT,''
  JHEP {\bf 1703}, 114 (2017)
  doi:10.1007/JHEP03(2017)114
  [arXiv:1702.01320 [hep-th]].
  %%CITATION = doi:10.1007/JHEP03(2017)114;%%
  %2 citations counted in INSPIRE as of 23 May 2017

%\cite{Buchel:2012gw}
\bibitem{Buchel:2012gw} 
  A.~Buchel, L.~Lehner and R.~C.~Myers,
  ``Thermal quenches in N=2* plasmas,''
  JHEP {\bf 1208}, 049 (2012)
  doi:10.1007/JHEP08(2012)049
  [arXiv:1206.6785 [hep-th]].
  %%CITATION = doi:10.1007/JHEP08(2012)049;%%
  %63 citations counted in INSPIRE as of 23 May 2017

%\cite{Buchel:2012uh}
\bibitem{Buchel:2012uh} 
  A.~Buchel, L.~Lehner and S.~L.~Liebling,
  ``Scalar Collapse in AdS,''
  Phys.\ Rev.\ D {\bf 86}, 123011 (2012)
  doi:10.1103/PhysRevD.86.123011
  [arXiv:1210.0890 [gr-qc]].
  %%CITATION = doi:10.1103/PhysRevD.86.123011;%%
  %76 citations counted in INSPIRE as of 23 May 2017


%\cite{Gursoy:2016ggq}
\bibitem{Gursoy:2016ggq} 
  U.~Gürsoy, A.~Jansen and W.~van der Schee,
  ``New dynamical instability in asymptotically anti–de Sitter spacetime,''
  Phys.\ Rev.\ D {\bf 94}, no. 6, 061901 (2016)
  doi:10.1103/PhysRevD.94.061901
  [arXiv:1603.07724 [hep-th]].
  %%CITATION = doi:10.1103/PhysRevD.94.061901;%%
  %13 citations counted in INSPIRE as of 23 May 2017

\end{thebibliography}
\end{document}